\documentclass[superscriptaddress,aps,prc,showkeys,showpacs,twocolumn,10pt]{revtex4-1}

\usepackage{amssymb}
\usepackage{amsmath}
\usepackage{graphicx}

\usepackage{subfigure}
\usepackage{paralist}
\usepackage{siunitx}

\begin{document}

\title{Measurements of Branching Ratios for $\eta$ Decays into Charged Particles}

\newcommand*{\IKPUU}{Division of Nuclear Physics, Department of Physics and 
 Astronomy, Uppsala University, Box 516, 75120 Uppsala, Sweden}
\newcommand*{\ASWarsN}{Department of Nuclear Physics, National Centre for 
 Nuclear Research, ul.\ Hoza~69, 00-681, Warsaw, Poland}
\newcommand*{\IPJ}{Institute of Physics, Jagiellonian University, ul.\ 
 Reymonta~4, 30-059 Krak\'{o}w, Poland}
\newcommand*{\PITue}{Physikalisches Institut, Eberhard--Karls--Universit\"at 
 T\"ubingen, Auf der Morgenstelle~14, 72076 T\"ubingen, Germany}
\newcommand*{\Kepler}{Kepler Center f\"ur Astro-- und Teilchenphysik, 
 Physikalisches Institut der Universit\"at T\"ubingen, Auf der 
 Morgenstelle~14, 72076 T\"ubingen, Germany}
\newcommand*{\Edinb}{School of Physics and Astronomy, University of Edinburgh, 
James Clerk Maxwell Building, Peter Guthrie Tait Road, Edinburgh EH9 3FD, 
Great Britain}
\newcommand*{\MS}{Institut f\"ur Kernphysik, Westf\"alische 
 Wilhelms--Universit\"at M\"unster, Wilhelm--Klemm--Str.~9, 48149 M\"unster, 
 Germany}
\newcommand*{\ASWarsH}{High Energy Physics Department, National Centre for 
 Nuclear Research, ul.\ Hoza~69, 00-681, Warsaw, Poland}
\newcommand*{\IITB}{Department of Physics, Indian Institute of Technology 
 Bombay, Powai, Mumbai--400076, Maharashtra, India}
\newcommand*{\Budker}{Budker Institute of Nuclear Physics of SB RAS, 
 11~akademika Lavrentieva prospect, Novosibirsk, 630090, Russia}
\newcommand*{\Novosib}{Novosibirsk State University, 2~Pirogova Str., 
 Novosibirsk, 630090, Russia}
\newcommand*{\IKPJ}{Institut f\"ur Kernphysik, Forschungszentrum J\"ulich, 
 52425 J\"ulich, Germany}
\newcommand*{\Bochum}{Institut f\"ur Experimentalphysik I, Ruhr--Universit\"at 
 Bochum, Universit\"atsstr.~150, 44780 Bochum, Germany}
\newcommand*{\ZELJ}{Zentralinstitut f\"ur Engineering, Elektronik und 
 Analytik, Forschungszentrum J\"ulich, 52425 J\"ulich, Germany}
\newcommand*{\Erl}{Physikalisches Institut, 
 Friedrich--Alexander--Universit\"at Erlangen--N\"urnberg, 
 Erwin--Rommel-Str.~1, 91058 Erlangen, Germany}
\newcommand*{\ITEP}{Institute for Theoretical and Experimental Physics, State 
 Scientific Center of the Russian Federation, 25~Bolshaya Cheremushkinskaya, 
 Moscow, 117218, Russia}
\newcommand*{\Giess}{II.\ Physikalisches Institut, 
 Justus--Liebig--Universit\"at Gie{\ss}en, Heinrich--Buff--Ring~16, 35392 
 Giessen, Germany}
\newcommand*{\IITI}{Department of Physics, Indian Institute of Technology 
 Indore, Khandwa Road, Indore--452017, Madhya Pradesh, India}
\newcommand*{\HepGat}{High Energy Physics Division, Petersburg Nuclear Physics 
 Institute, 2~Orlova Rosha, Gatchina, Leningrad district, 188300, Russia}
\newcommand*{\HeJINR}{Veksler and Baldin Laboratory of High Energiy Physics, 
 Joint Institute for Nuclear Physics, 6~Joliot--Curie, Dubna, 141980, Russia}
\newcommand*{\Katow}{August Che{\l}kowski Institute of Physics, University of 
 Silesia, Uniwersytecka~4, 40-007, Katowice, Poland}
\newcommand*{\IFJ}{The Henryk Niewodnicza{\'n}ski Institute of Nuclear 
 Physics, Polish Academy of Sciences, 152~Radzikowskiego St, 31-342 
 Krak\'{o}w, Poland}
\newcommand*{\JARA}{JARA--FAME, J\"ulich Aachen Research Alliance, 
 Forschungszentrum J\"ulich, 52425 J\"ulich, and RWTH Aachen, 52056 Aachen, 
 Germany}
\newcommand*{\Tomsk}{Department of Physics, Tomsk State University, 36~Lenina 
 Avenue, Tomsk, 634050, Russia}
\newcommand*{\KEK}{High Energy Accelerator Research Organisation KEK, Tsukuba, 
 Ibaraki 305--0801, Japan} 
\newcommand*{\ASLodz}{Department of Astrophysics, National Centre for Nuclear 
 Research, ul.\ Box 447, 90--950 {\L}\'{o}d\'{z}, Poland}

\author{P.~Adlarson}\altaffiliation[present address: ]{\Mainz}\affiliation{\IKPUU}
\author{W.~Augustyniak} \affiliation{\ASWarsN}
\author{W.~Bardan}      \affiliation{\IPJ}
\author{M.~Bashkanov}\altaffiliation[present address: ]{\Edinb}\affiliation{\PITue}\affiliation{\Kepler}
\author{F.S.~Bergmann}  \affiliation{\MS}
\author{M.~Ber{\l}owski}\affiliation{\ASWarsH}
\author{H.~Bhatt}       \affiliation{\IITB}
\author{A.~Bondar}      \affiliation{\Budker}\affiliation{\Novosib}
\author{M.~B\"uscher}\altaffiliation[present address: ]{\PGI, \DUS}\affiliation{\IKPJ}
\author{H.~Cal\'{e}n}   \affiliation{\IKPUU}
\author{I.~Ciepa{\l}}   \affiliation{\IPJ}
\author{H.~Clement}     \affiliation{\PITue}\affiliation{\Kepler}   
\author{D.~Coderre}  \email[corresponding author email: ]{daniel.coderre@lhep.unibe.ch} \altaffiliation[present address: ]{\Bern}
\affiliation{\IKPJ}\affiliation{\Bochum}
\author{E.~Czerwi{\'n}ski}\affiliation{\IPJ}
\author{K.~Demmich}     \affiliation{\MS}
\author{R.~Engels}      \affiliation{\IKPJ}
\author{A.~Erven}       \affiliation{\ZELJ}
\author{W.~Erven}       \affiliation{\ZELJ}
\author{W.~Eyrich}      \affiliation{\Erl}
\author{P.~Fedorets}    \affiliation{\IKPJ}\affiliation{\ITEP}
\author{K.~F\"ohl}      \affiliation{\Giess}
\author{K.~Fransson}    \affiliation{\IKPUU}
\author{F.~Goldenbaum}  \affiliation{\IKPJ}
\author{A.~Goswami}     \affiliation{\IKPJ}\affiliation{\IITI}
\author{K.~Grigoryev}\altaffiliation[present address: ]{\Aachen}\affiliation{\IKPJ}\affiliation{\HepGat}
\author{C.--O.~Gullstr\"om}\affiliation{\IKPUU}
\author{L.~Heijkenskj\"old}\affiliation{\IKPUU}
\author{V.~Hejny}       \affiliation{\IKPJ}
\author{N.~H\"usken}    \affiliation{\MS}
\author{L.~Jarczyk}     \affiliation{\IPJ}
\author{T.~Johansson}   \affiliation{\IKPUU}
\author{B.~Kamys}       \affiliation{\IPJ}
\author{G.~Kemmerling}  \affiliation{\ZELJ}
\author{F.A.~Khan}      \affiliation{\IKPJ}
\author{G.~Khatri}      \affiliation{\IPJ}
\author{A.~Khoukaz}     \affiliation{\MS}
\author{D.A.~Kirillov}  \affiliation{\HeJINR}
\author{S.~Kistryn}     \affiliation{\IPJ}
\author{H.~Kleines}     \affiliation{\ZELJ}
\author{B.~K{\l}os}     \affiliation{\Katow}
\author{W.~Krzemie{\'n}}\affiliation{\IPJ}
\author{P.~Kulessa}     \affiliation{\IFJ}
\author{A.~Kup\'{s}\'{c}}\affiliation{\IKPUU}\affiliation{\ASWarsH}
\author{A.~Kuzmin}       \affiliation{\Budker}\affiliation{\Novosib}
\author{K.~Lalwani}\altaffiliation[present address: ]{\Delhi}\affiliation{\IITB}
\author{D.~Lersch}      \affiliation{\IKPJ}
\author{B.~Lorentz}     \affiliation{\IKPJ}
\author{A.~Magiera}     \affiliation{\IPJ}
\author{R.~Maier}       \affiliation{\IKPJ}\affiliation{\JARA}
\author{P.~Marciniewski}\affiliation{\IKPUU}
\author{B.~Maria{\'n}ski}\affiliation{\ASWarsN}
\author{M.~Mikirtychiants}\affiliation{\IKPJ}\affiliation{\Bochum}\affiliation{\HepGat}
\author{H.--P.~Morsch}  \affiliation{\ASWarsN}
\author{P.~Moskal}      \affiliation{\IPJ}
\author{H.~Ohm}         \affiliation{\IKPJ}
\author{I.~Ozerianska}  \affiliation{\IPJ}
\author{E.~Perez del Rio}\altaffiliation[present address: ]{\INFN}\affiliation{\PITue}\affiliation{\Kepler}
\author{N.M.~Piskunov}  \affiliation{\HeJINR}
\author{D.~Prasuhn}     \affiliation{\IKPJ}
\author{D.~Pszczel}     \affiliation{\IKPUU}\affiliation{\ASWarsH}
\author{K.~Pysz}        \affiliation{\IFJ}
\author{A.~Pyszniak}    \affiliation{\IKPUU}\affiliation{\IPJ}
\author{J.~Ritman}\affiliation{\IKPJ}\affiliation{\Bochum}\affiliation{\JARA}
\author{A.~Roy}         \affiliation{\IITI}
\author{Z.~Rudy}        \affiliation{\IPJ}
\author{O.~Rundel}      \affiliation{\IPJ}
\author{S.~Sawant}      \affiliation{\IITB}\affiliation{\IKPJ}
\author{S.~Schadmand}   \affiliation{\IKPJ}
\author{T.~Sefzick}     \affiliation{\IKPJ}
\author{V.~Serdyuk}     \affiliation{\IKPJ}
\author{B.~Shwartz}     \affiliation{\Budker}\affiliation{\Novosib}
\author{K.~Sitterberg}  \affiliation{\MS}
\author{R.~Siudak}      \affiliation{\IFJ}
\author{T.~Skorodko}\affiliation{\PITue}\affiliation{\Kepler}\affiliation{\Tomsk}
\author{M.~Skurzok}     \affiliation{\IPJ}
\author{J.~Smyrski}     \affiliation{\IPJ}
\author{V.~Sopov}       \affiliation{\ITEP}
\author{R.~Stassen}     \affiliation{\IKPJ}
\author{J.~Stepaniak}   \affiliation{\ASWarsH}
\author{E.~Stephan}     \affiliation{\Katow}
\author{G.~Sterzenbach} \affiliation{\IKPJ}
\author{H.~Stockhorst}  \affiliation{\IKPJ}
\author{H.~Str\"oher}   \affiliation{\IKPJ}\affiliation{\JARA}
\author{A.~Szczurek}    \affiliation{\IFJ}
\author{A.~T\"aschner}  \affiliation{\MS}
\author{A.~Trzci{\'n}ski}\affiliation{\ASWarsN}
\author{R.~Varma}       \affiliation{\IITB}
\author{U.~Wiedner}     \affiliation{\Bochum}
\author{M.~Wolke}       \affiliation{\IKPUU}
\author{A.~Wro{\'n}ska} \affiliation{\IPJ}
\author{P.~W\"ustner}   \affiliation{\ZELJ}
\author{P.~Wurm}        \affiliation{\IKPJ}
\author{A.~Yamamoto}    \affiliation{\KEK}
\author{J.~Zabierowski} \affiliation{\ASLodz}
\author{M.J.~Zieli{\'n}ski}\affiliation{\IPJ}
\author{A.~Zink}        \affiliation{\Erl}
\author{J.~Z{\l}oma{\'n}czuk}\affiliation{\IKPUU}
\author{P.~{\.Z}upra{\'n}ski}\affiliation{\ASWarsN}
\author{M.~{\.Z}urek}   \affiliation{\IKPJ}

\newcommand*{\Mainz}{Institut f\"ur Kernphysik, Johannes 
 Gutenberg--Universit\"at Mainz, Johann--Joachim--Becher Weg~45, 55128 Mainz, 
 Germany}
\newcommand*{\PGI}{Peter Gr\"unberg Institut, PGI--6 Elektronische 
 Eigenschaften, Forschungszentrum J\"ulich, 52425 J\"ulich, Germany}
\newcommand*{\DUS}{Institut f\"ur Laser-- und Plasmaphysik, Heinrich--Heine 
 Universit\"at D\"usseldorf, Universit\"atsstr.~1, 40225 D\"usseldorf, Germany}
\newcommand*{\Bern}{Albert Einstein Center for Fundamental Physics, 
 Universit\"at Bern, Sidlerstrasse~5, 3012 Bern, Switzerland}
\newcommand*{\Aachen}{III.~Physikalisches Institut~B, Physikzentrum, 
 RWTH Aachen, 52056 Aachen, Germany}
\newcommand*{\Delhi}{Department of Physics and Astrophysics, University of 
 Delhi, Delhi--110007, India}
\newcommand*{\INFN}{INFN, Laboratori Nazionali di Frascati, Via E. Fermi, 40, 
 00044 Frascati (Roma), Italy}

\collaboration{WASA-at-COSY Collaboration}\noaffiliation


\begin{abstract}
The WASA-at-COSY experiment has collected $3\times10^{7}$
events with $\eta$-mesons produced via the reaction
$pd\rightarrow{^{3}\textrm{He}}\eta$ at $\textrm{T} = 1.0
\textrm{GeV}$. Using this data set, we evaluate the branching ratios of
the decays $\eta\to\pi^{+}\pi^{-}\gamma$, $\eta\to e^{+}e^{-}\gamma$,
$\eta\to\pi^{+}\pi^{-}e^{+}e^{-}$ and $\eta\to e^{+}e^{-}e^{+}e^{-}$.
The branching ratios are normalized to the
$\eta\rightarrow\pi^{+}\pi^{-}\pi^{0}$ decay. In addition an upper
limit on a $CP$-violating asymmetry in $\eta\rightarrow
\pi^{+}\pi^{-}e^{+}e^{-}$ is extracted.
\end{abstract}

\maketitle





\section{Introduction}
\label{1_Intro}

Studies of the strong interaction at low energies are vital to the
understanding of the structure and dynamics of hadrons as well as the
nature of confinement.  At low energies, the quantum chromodynamics
(QCD) coupling becomes large and standard perturbative methods cannot
be used.  The main theoretical approaches at low energies are lattice
QCD and effective field theories, including chiral perturbation
theory.  Precise measurements at these energies provide
valuable inputs and can constrain and test these approaches.

The $\eta$ meson is one of the eight pseudo-Goldstone bosons of the
broken chiral symmetry and therefore studies of its decays provide a
unique window into low-energy QCD. The $\eta$ meson is a light, neutral pseudoscalar 
with a mass of 
$(547.862\pm0.018)$ $\textrm{MeV/c}^{2}$ \cite{PDG}.  All strong and electromagnetic decays of the $\eta$ are
forbidden to first order, resulting in a relatively long lifetime
and a correspondingly narrow width of $1.31 \pm 0.05$ keV. This makes
the $\eta$ meson an ideal laboratory for the study of rare processes,
since the suppression of many of the more abundant decay modes makes
rare decays experimentally accessible.

We  report the  measurement of the branching  ratios of  the following
four $\eta$ meson decay channels:
\begin{eqnarray*}
\eta&\rightarrow&\pi^{+}\pi^{-}\gamma,\\
\eta&\rightarrow&e^{+}e^{-}\gamma,\\
\eta&\rightarrow&\pi^{+}\pi^{-}e^{+}e^{-},\\  
\eta&\rightarrow&  e^{+}e^{-}e^{+}e^{-}
\end{eqnarray*}
  collected in proton-deuteron collisions at the WASA-at-COSY
  experiment using the $\eta\rightarrow\pi^{+}\pi^{-}\pi^{0}$ decay with
  $\pi^{0}\to\gamma\gamma(\equiv\pi^{0}_{\gamma\gamma})$ as the
  normalization channel.

Using a minimum bias data sample of $\eta$ mesons and the
reconstruction capabilities of the WASA detector, most notably the
charged particle tracking and particle identification, we are able to
isolate pure samples of several decay modes.  It is important to note
that these are the only current results on $\eta$ decays where the
$\eta$-mesons are produced in hadronic interactions, therefore they
feature complementary experimental conditions compared to the results
of other experiments which use photoproduction or $e^{+}e^{-}$
collisions for meson production.

\section{The Experiment}
\label{2_WASA}
The WASA-at-COSY experiment was operated at the Cooler
Synchrotron (COSY) at Forschungszentum J\"ulich from 2006 to 2014
\cite{Hoistad2004}. For the data used in this analysis, a proton beam
with $\textrm{T}=1.0\, \textrm{GeV}$ was impinged upon a deuterium
pellet target. The reaction ${pd}\rightarrow {^{3}\textrm{He}}\,\eta$ is
used to produce $\eta$ mesons at energies close to the production
threshold, where the most favorable ratio between the $\eta$ production cross section 
and background reactions is found. The cross section of this production reaction is $0.40(3) \mu\textrm{b}$
\cite{Bilger2002,PhysRevC.80.017001}, meaning that up to 8 events containing $\eta$ mesons
are produced per second at the peak luminosity of $2\times10^{31}$
cm$^{-2}$s$^{-1}$.

\begin{figure*}[htb]
\centering \includegraphics[width=0.9\textwidth]{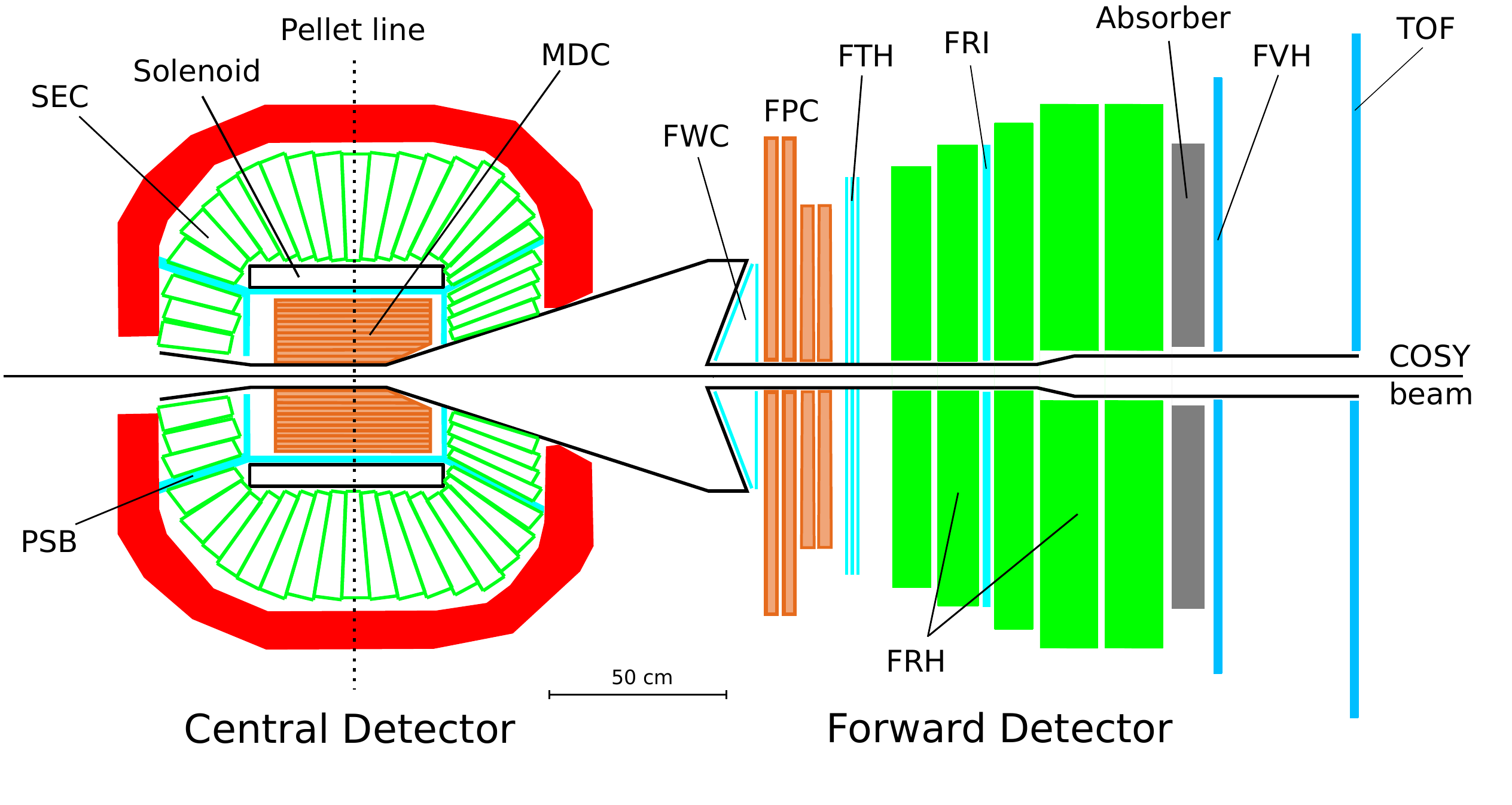}
  \caption[The WASA Detector]{(Color online) A cross-sectional scheme of
    the WASA detector with the beam coming from the
    left. Hadronic ejectiles are measured with the forward detector on
    the right while meson decay products are measured with the central
    detector on the left. Components are described in the
    text.}
 \label{fig:wasadet}
\end{figure*}

The WASA detector is a fixed-target
spectrometer, with a forward detector arranged to measure hadronic
ejectiles and a central detector to detect light mesons or their decay
products. A cross-sectional view of the detector appears in
Fig.~\ref{fig:wasadet}. The forward detector consists of an
arrangement of thin and thick plastic scintillators and drift
chambers covering the full azimuthal angle.  Thick scintillators in the
forward range hodoscope (FRH) are designed to measure energy loss via
ionization. Thin scintillator layers in the forward window counter
(FWC) and forward trigger hodoscope (FTH) provide precise timing
information. The kinetic energy and the particle type can be
determined from the pattern of energy deposits in the thin and thick
scintillator layers.  A proportional chamber system (FPC) consists of
8 layers, each with 260 aluminized Mylar straws. Layers of the forward detector 
beyond the first layer of the FRH, including the Forward Range Interleaving Hodoscope (FRI) 
detector and the Forward Veto Hodoscope (FVH), were not used in this analysis 
due to the kinematics of the reaction.

The central detector is surrounded by a CsI(Na) electromagnetic calorimeter with 1012
elements (SEC).  Contained within the calorimeter is a superconducting solenoid 
providing a uniform 1 T magnetic field to the region directly surrounding the interaction area.
Charged particle tracking in this region is provided by the mini drift chamber (MDC), which is surrounded by an 8 mm thick plastic scintillator barrel
(PSB) that provides precise timing and particle identification. 
The MDC consists of 4, 6, and 8 mm diameter straw tubes
arranged in 17 layers that are alternatingly axial or skewed by
$+3^\circ$ or $-3^\circ$ relative to the beam axis in order to provide three-dimensional
tracking. An iron return yoke, shown in red in Fig.~\ref{fig:wasadet}, surrounds 
the central detector and protects the photomultplier tubes of the SEC from the magnetic field. A
detailed description of the WASA detector can be found in
Ref.~\cite{Bargholtz2008,Hoistad2004}.

The data for this experiment were taken over two periods with
four weeks in the fall of 2008 and eight weeks in the fall of 2009.
Care was taken to provide consistent conditions between the two run
periods.  The solenoid field setting was 0.85 Tesla.  The
trigger conditions were based on information from the forward detector
only, meaning the trigger was unbiased with respect to a
decay mode of the $\eta$ meson.  The trigger identified
${^{3}\textrm{He}}$ ions by demanding large energy deposits in
overlapping azimuthal sectors of the scintillator layers.  In the case
of the ${pd}\rightarrow {^{3}\textrm{He}}\,\eta$ reaction the
${^{3}\textrm{He}}$ stops in the first layer of FRH, so the
trigger included a veto on the signals from the second layer.  Since
fusion to a ${{^3}\textrm{He}}$ represents only about 1\% of the total
cross section at this energy, the above conditions were sufficient to bring the 
trigger rate down to below a thousand events per second, which were recorded.  
$3\times10^{7}$ events containing $\eta$ mesons
were collected in total, with $1\times10^{7}$ being collected during
the first period in 2008.

\section{Event Selection}
\label{3_Analysis}
\begin{figure}[!h]
\begin{center}
 \includegraphics[width=0.45\textwidth]{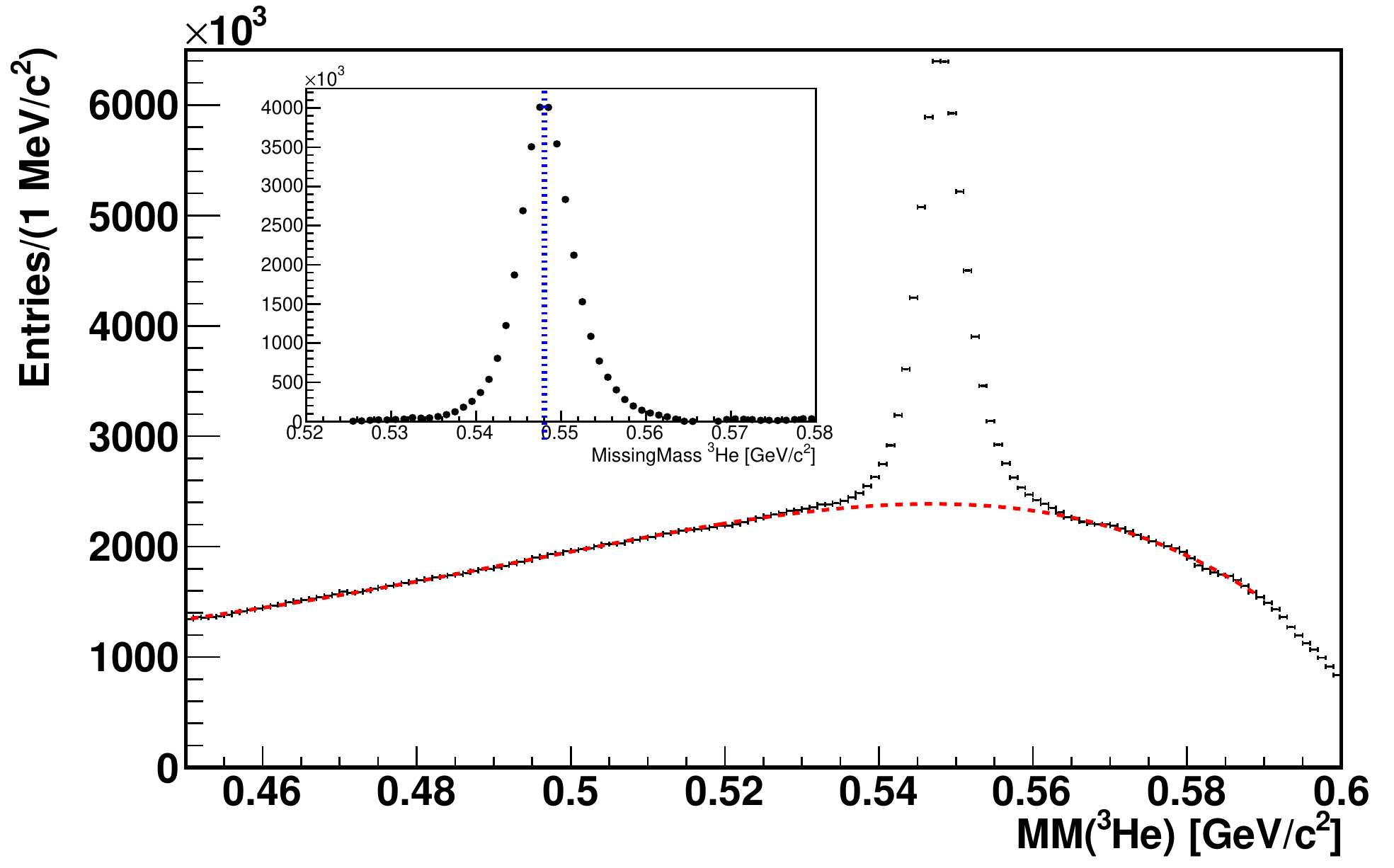}
 \caption[Unbiased Missing Mass Peak]{(Color online) Initial spectrum of the ${^{3}\textrm{He}}$ missing mass. 
 The prominent peak is due to $\eta$ meson production, while the broad background distribution is from production of 
 two and three $\pi$-mesons. The dashed (red) line shows a fit of the background (incorporating the shapes of double 
 and triple $\pi$ meson production, as described in the text) and the inset shows the peak after background 
 subtraction with the $\eta$ mass marked by the dotted (blue) line.}
\label{fig:mmhe3}
\end{center}
\end{figure}

All decay channels are analyzed using a common analysis chain and settings
up to the point of channel selection and kinematic fitting.  The first
step is to identify the ${^{3}\textrm{He}}$ ion. Forward-scattered
tracks are reconstructed by using hit patterns in the FPC and matching
them to signals in the scintillator layers.  To separate
${^{3}\textrm{He}}$ ions from protons, deuterons, and charged
$\pi$-mesons, the energy deposited in the FWC is correlated with the
energy deposited in the stopping layer. Once the ${^{3}\textrm{He}}$ ion is identified, 
the missing mass, MM$({^{3}\textrm{He}})$,
can be calculated by determining the invariant mass remaining when the measured ${^{3}\textrm{He}}$
four vector is subtracted from the known initial conditions of the beam and target.  
The resulting distribution is shown in
Fig.~\ref{fig:mmhe3} with a peak at the mass of the $\eta$ meson. This peak is 
composed of all decay modes of the $\eta$, since this stage of the analysis does not include 
any condition on the central detector.  This initial sample contains $3\times10^{7}$ events
with $\eta$ mesons.

Particles from $\eta$ meson decays are measured in the central
detector.  Tracks reconstructed in the MDC are extrapolated to the PSB
and to the calorimeter. Clusters in the calorimeter that are not
correlated with the tracks are treated as electromagnetic showers caused by photons. A
threshold is placed on a cluster energy of 20 MeV to filter out low-energy background signals. The PSB time
signals with a time resolution of  $1$ ns are used to
provide a start signal for the drift-time measurement  in the MDC.  All
tracks are required to pass closer than 1 cm from the beam axis. 

Event candidates for the $\eta$ decay channels have to include a
${^{3}\textrm{He}}$ ion in time coincidence with at least the minimum
number of tracks (charged particles) and neutral clusters (photons) for a selected decay mode.

Energy and momentum conservation is imposed by requiring all event
candidates to pass through a kinematic fitting routine for a specific
$\eta$ decay channel.  The kinematic fit takes into account
reconstruction uncertainties for the different particle types as a
function of angles and energies. A condition on the mass of the
decaying $\eta$ or $\pi^{0}$ meson is not imposed.  Events with a fit
probability less than 0.1 are rejected.  For certain channels
additional conditions are applied, which are described in the later
sections.

\section{Simulations}

Efficiencies used for acceptance correction are calculated using Monte
Carlo simulations.  Kinematic event  generation is performed  with the
PLUTO$++$  software  package  \cite{pluto}.  This  contains  realistic
physical descriptions  of all relevant channels. The
angular  distribution  of  the  produced  $\eta$  mesons  measured  in
Ref. \cite{WASA_etadiff} for the $pd\rightarrow {^{3}\textrm{He}}\,\eta$
production reaction is used.

For $\eta\rightarrow\pi^{+}\pi^{-}\pi^{0}$, the Dalitz plot parameters
from the Crystal Barrel \cite{Amsler1995} measurement are used.  The
simulation of channels $\eta\rightarrow\pi^{+}\pi^{-}\gamma^{(*)}$ is
based on calculations from Ref.~\cite{Petri2010}.  The decays
$\eta\rightarrow e^{+}e^{-}\gamma^{(*)}$ are simulated using form
factors calculated assuming the vector meson dominance model with
the transition form factor $F(q^2)=1/(1-b_\eta q^2)$ where $q^2$ is the
invariant mass squared of the electron pair and parameter $b_\eta=1.78
 \textrm{ GeV}^{-2}$ \cite{Landsberg1985}.

Detector simulations are performed using the WASA Monto Carlo package,
WMC, which is based on GEANT3 \cite{geant}.  Temporal, spatial, and
energy resolution of the detector elements is implemented in the WMC
using data to fine-tune the parameters.  For example, the
$\pi^{0}\rightarrow\gamma\gamma$ and $\eta\rightarrow\gamma\gamma$
decays are used to determine the energy resolution of the calorimeter
by analyzing the two photon invariant mass
distributions. Additionally, inactive detector channels are
continuously monitored and mapped in the simulations. The number of inactive channels is typically less than a few percent.

The WASA-at-COSY experiment uses an internal target with frozen
pellets injected at rates of several thousand per second
\cite{Ekstroem1996}. Though vacuum pumps are positioned as closely as
possible to the interaction region, a certain amount of residual gas is
present in the region around the target which comes from the
evaporation of pellets.  This is quantified in data by selecting
$pd\rightarrow {^{3}\textrm{He}}\,\pi^{+}\pi^{-}$ events and
reconstructing the vertex from the $\pi^{+}\pi^{-}$ tracks. The resulting
spectrum has a large spike in the target region (with dimensions
determined by the profiles of the beam and the pellet stream) as well
as tails along the beam axis due to beam-gas interactions. Monte Carlo
simulations of these ``rest gas'' events were performed by including
the shape of the vertex distribution in the $z$ direction deduced from
the experimental data.  Over 90\% of all events occur within one
centimeter of the center of the interaction region.

\section{Particle Identification}
\label{sec_pid}
Three of the final states being studied contain electrons and
positrons in the final state. In
WASA-at-COSY, above the kinematic threshold for double $\pi^{0}\pi^{0}$ and 
$\pi^{+}\pi^{-}$ production the dominant background contributions stem from $\pi$ meson production.
In the case of $\eta\rightarrow
e^{+}e^{-}\gamma$ and $\eta\rightarrow e^{+}e^{-}e^{+}e^{-}$, the
final states are mimicked by more abundant channels containing charged
$\pi$-mesons.  In the case of
$\eta\rightarrow\pi^{+}\pi^{-}e^{+}e^{-}$, pions have to be identified
in order to reconstruct the kinematics of the final state.  For the two
purposes ($\pi$-meson rejection {\it vs.}  identification) two
slightly different algorithms are used. %

\begin{figure}[!h] 
\begin{center}
\includegraphics[width=0.45\textwidth]{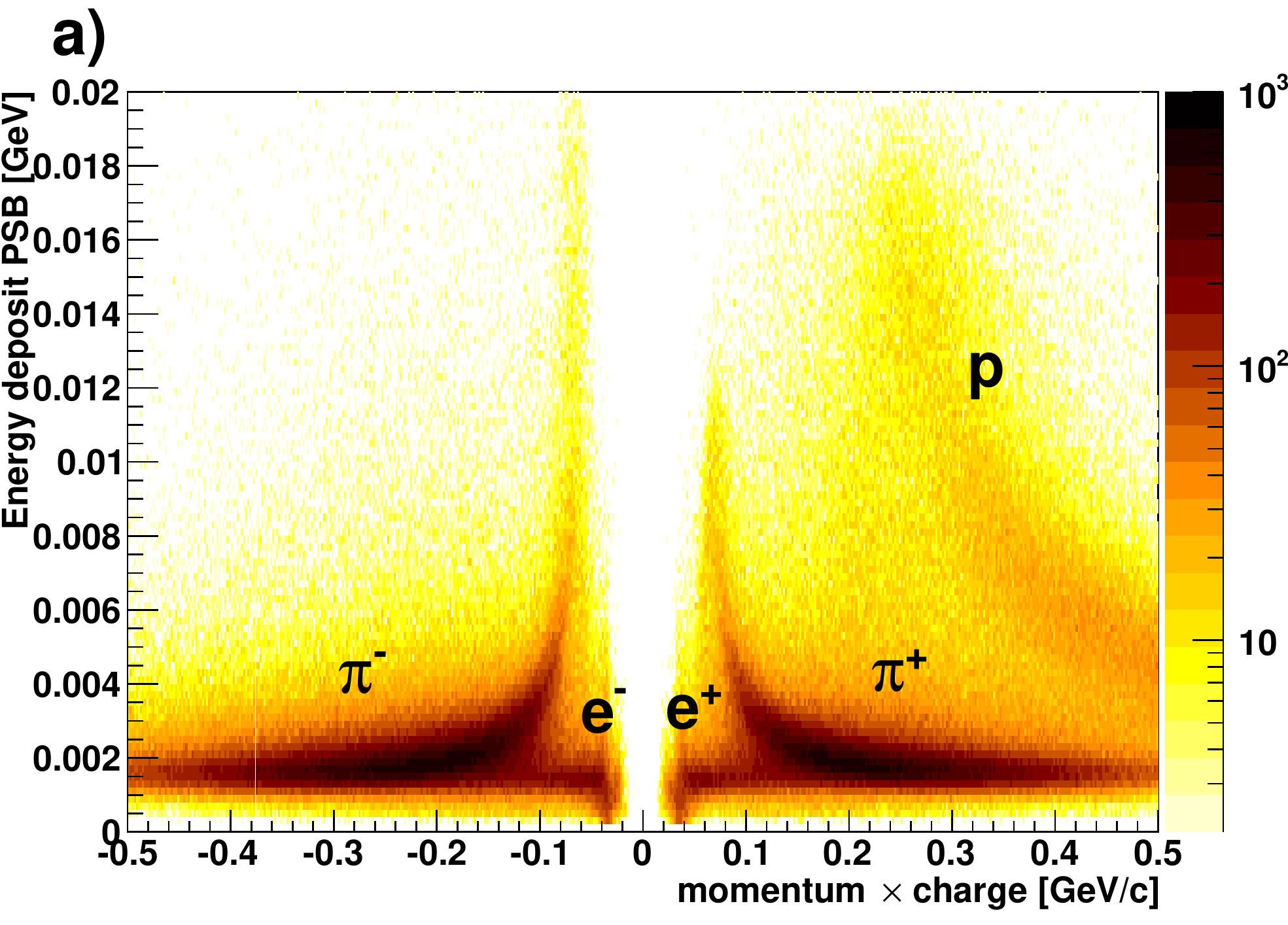}
\includegraphics[width=0.45\textwidth]{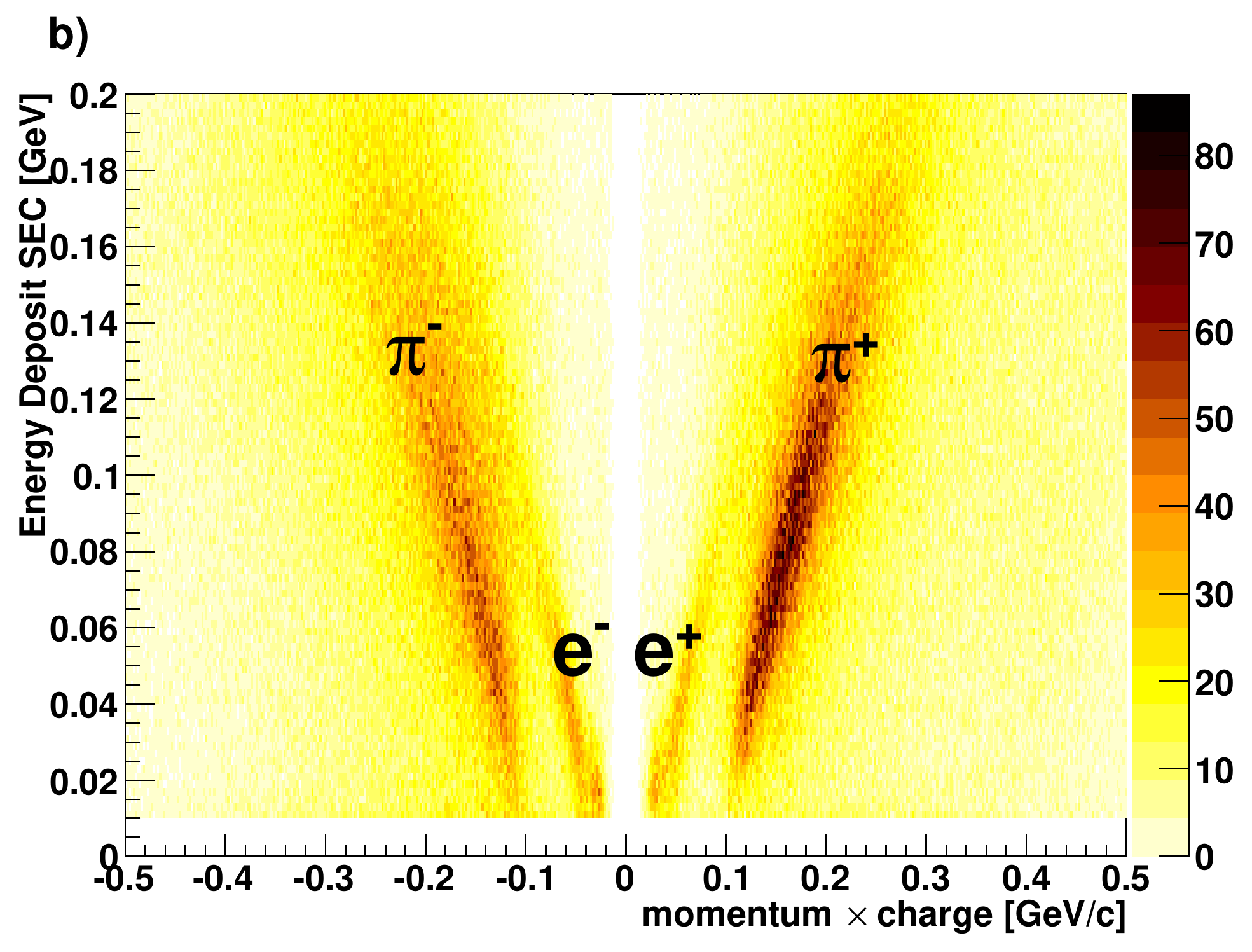}
\caption[Particle Identification Bands]{(Color online) Charge-signed
  momentum versus energy deposit in the a) plastic scintillator barrel (PSB)
  and b) calorimeter (SEC). The bands corresponding to $e^{+/-}$ and $\pi$-mesons are labeled. }
\label{fig:pid}
\end{center}
\end{figure}

The WASA  detector provides an independent  measurement of the
momentum of a  charged particle in the MDC, as well as
energy loss   in  the   plastic   scintillator  barrel   and  in   the
electromagnetic   calorimeter.      Fig.~\ref{fig:pid}   shows   the
distributions of the deposited  energy versus  the momentum  times the
particle's  charge for energies measured in both the PSB and the calorimeter.  
The energy  loss in  the plastic  scintillator is
corrected  for  the track  length  of  the  particle in  the  detector
element.   In order  to illustrate  the discriminating  power  of this
method,  these  spectra were  created  for  events  with four  charged
particle  tracks,  since  a  large  number  of  these  events  contain
electrons.   The  bands  from  charged  $\pi$-mesons,  electrons,  and
positrons are labeled.

To  utilize this  information,  a simple  Bayesian  approach has  been
developed which allows the discrimination of $\pi$-mesons from $e^{+/-}$
using all pieces of information simultaneously \cite{CoderrePhD}.  Two algorithms exist:
a rejection  algorithm, which  considers two particles  at a  time and
aims at rejecting $\pi$-meson pairs, and  a classification algorithm,
which considers  four particles at a  time and aims at assigning particle types.

\begin{figure}[!h] 
\begin{center}
 \includegraphics[width=0.45\textwidth]{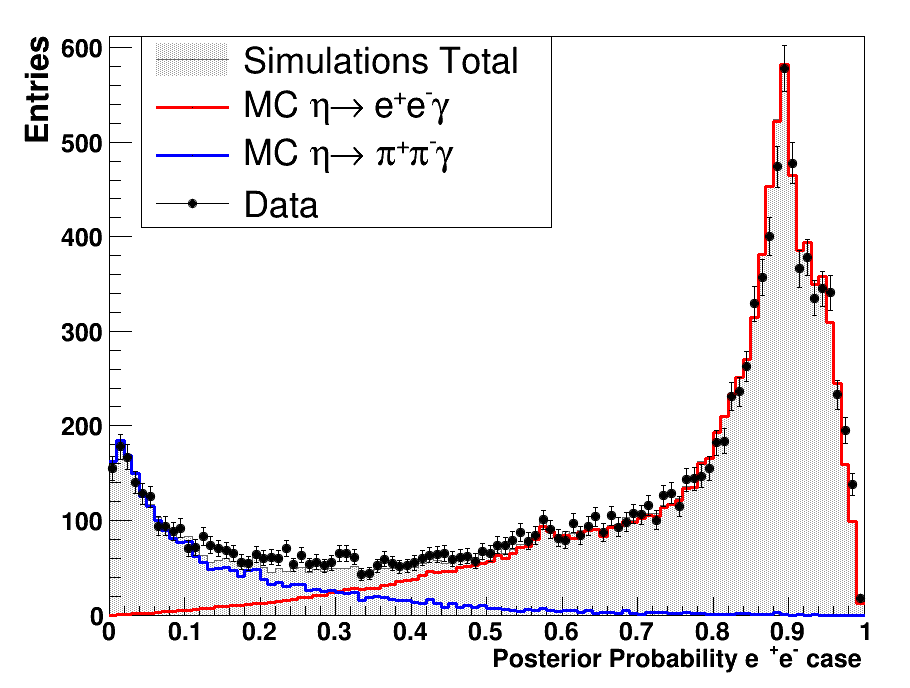}
 \caption[PID Performance]{(Color online) Posterior probabilities for events passing the $\eta\rightarrow e^{+}e^{-}\gamma$ kinematic fit hypothesis showing the discrimination of $\pi$-mesons from electrons. The red and blue lines are simulations of $\eta\rightarrow e^{+}e^{-}\gamma$ and $\eta\rightarrow\pi^{+}\pi^{-}\gamma$ respectively and the shaded area is the sum of both simulations.}
\label{fig:pidtestcocktail}
\end{center}
\end{figure}

For the rejection algorithm, pairs of oppositely-charged particles
are considered: a pair could be either an $e^{+}e^{-}$ or a
$\pi^{+}\pi^{-}$ with equal {\it a priori} probability. The
probability that a single particle is a $\pi$-meson or a lepton is
determined from the momentum and energy losses and added to the
posteriors using Bayes' equation.  After considering both particles,
the configuration with the highest probability is chosen. A
distribution of the posterior probabilities in data and simulation is
compared in Fig.~\ref{fig:pidtestcocktail}, where a clear separation
between particle types can be seen. The distribution is made for
events passing the $pd\to{^{3}\textrm{He}}\, e^{+}e^{-}\gamma$ kinematic fit
hypothesis with a $({^{3}\textrm{He}})$ missing mass within $\pm1$
$\textrm{MeV}/\textrm{c}^{2}$ of the actual $\eta$ meson mass,
in order to enhance the electron contribution.

The  graphical identification  bands  shown in
Fig.~\ref{fig:pid}  are represented as probabilities using neural networks  from the ROOT
{\texttt{ TMultilayerPerceptron}} class \cite{ROOT}. The neural networks
are  trained  using  simulated  $\pi$-mesons  and  electrons  tracks with  
isotropic directions and a flat 
 energy distribution as signal and an uncorrelated, randomly-generated
data  set as  background. The  likelihood  function that  is used  for
Bayes' equation is statistically determined from an independent set of
simulated  data. The  resolution  and position  of the  identification
bands  shown  in Fig.~\ref{fig:pid}  was  tuned  in simulations  to
describe the data before the neural networks were trained.

For the $\eta\rightarrow\pi^{+}\pi^{-}e^{+}e^{-}$ analysis a
classification algorithm is used, as described in Section
\ref{sec_etappee}.

\section{Photon Conversion Suppression}
\label{sec_conv}

Final state electron-positron pairs originate
from virtual photons where the corresponding radiative decay has a branching ratio 
that is about two orders of magnitude larger. The external conversion of
real photons is suppressed by the design of the WASA detector, which uses
a thin beryllium beam pipe in the interaction region. However, there is still a 1\%
chance that a photon will convert in the beam pipe or the inner
layers of the drift chamber, producing an electron-positron pair. This means the
magnitude of the conversion background is similar to that of the
signal.

\begin{figure}[!h] 
\begin{center}
\subfigure{
\includegraphics[width=0.48\textwidth]{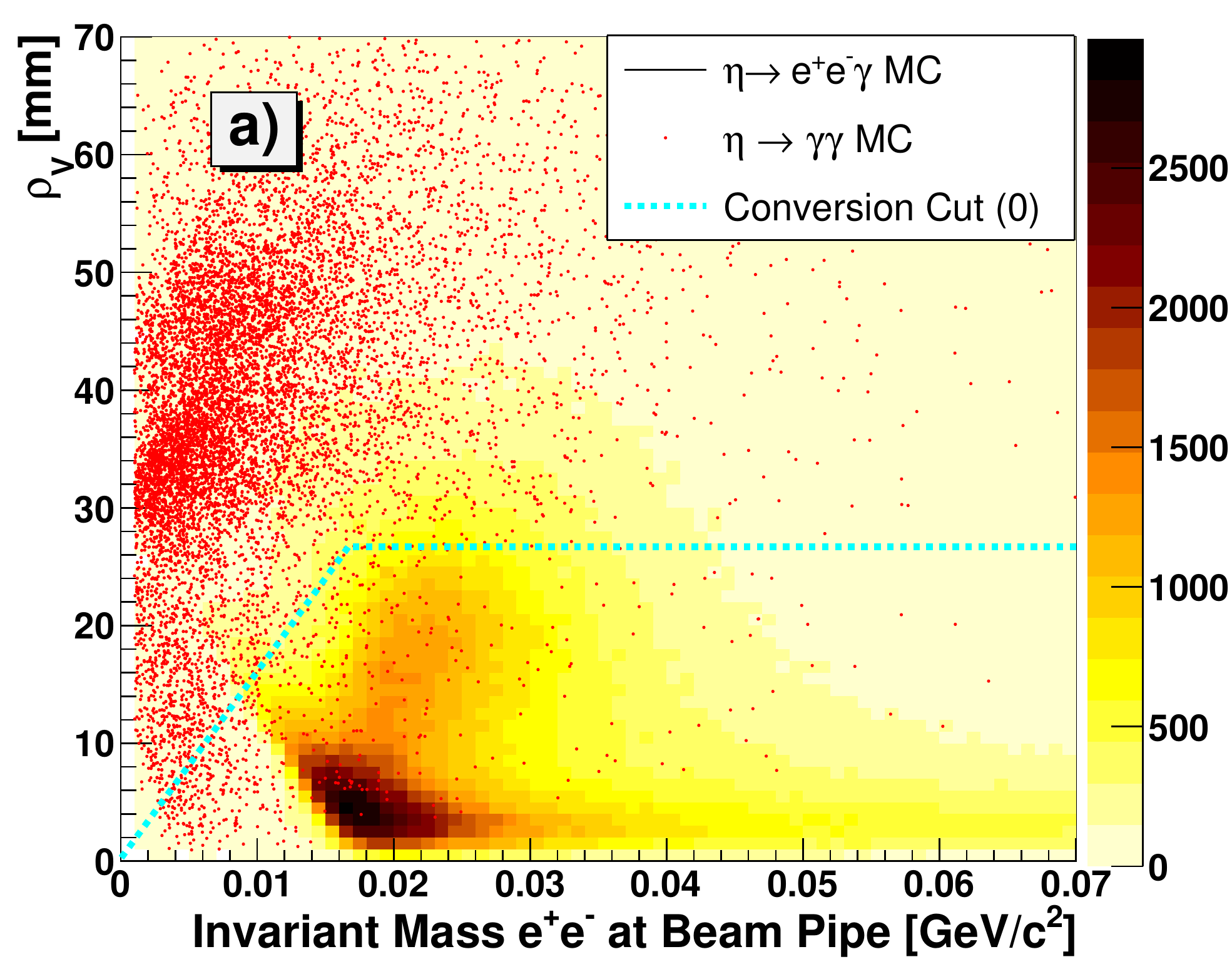}}
\subfigure{
 \includegraphics[width=0.48\textwidth]{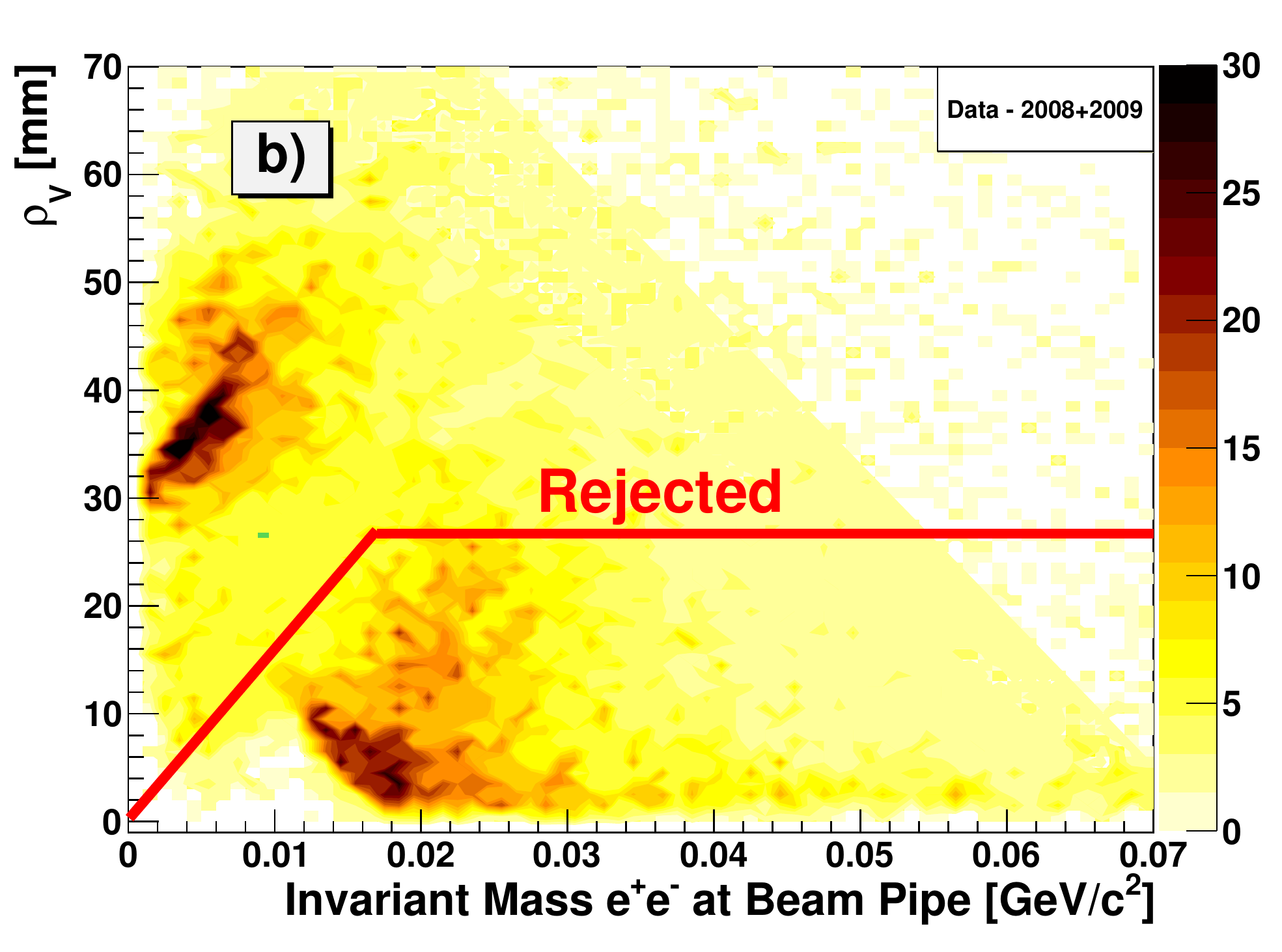}}
 \caption[Conversion Condition]{(Color online) Condition on conversion
   suppression shown for a) simulations and b) data. For the
   simulations, the multi-colored plot is for the $\eta\rightarrow
   e^{+}e^{-}\gamma$ signal while the red dots are for
   $\eta\rightarrow\gamma\gamma$ where a photon converts in the
   beam pipe. The demarcation lines shown in each plot are the same.}
\label{fig:conv}
\end{center}
\end{figure}

In order to suppress this background, the electron-positron vertex
position is determined from the reconstructed MDC tracks.  For events
where the electron-positron pair originates from the actual
$\eta$ meson decay, the vertex is close to the center of the
interaction region.  For events where the particles are the result of
photon conversion in the detector material, the vertex distance will
be at least equal to the radius of the beam pipe.  The vertex distance
in the plane perpendicular to the COSY beam, $\rho_V$, is represented
on the $y$-axis in Fig.~\ref{fig:conv} for simulations and data.  For
simulations the channel $\eta\rightarrow\gamma\gamma$ is shown in red
while the colored spectrum represents $\eta\rightarrow
e^{+}e^{-}\gamma$.  The $\eta\rightarrow\gamma\gamma$ events are
clustered starting at $\rho_V$ of about 30 mm, which corresponds to
the radius of the beam pipe.  The $\eta\rightarrow e^{+}e^{-}\gamma$
events are clustered around zero.

The invariant mass of the  $e^{+}e^{-}$ pair at the beam pipe location
provides  additional discriminating  power, as  inspired by  a similar
condition used  in Ref. \cite{Ambrosino2009}.  Normally, the direction
of the momentum vector of a particle is set to be tangent to the track
helix at the closest approach  to the origin. For this calculation the
momentum vector is  recalculated at the point where  the helix crosses
the  beam pipe.  This does not change  the magnitude  of the  momentum
vector, which is  determined by the radius of the  helix, or the polar
angle, which  is determined by  the pitch of  the helix, but  changes the
azimuthal  angle.  The expectation  is  that  the  momentum vectors  of
an electron-positron pair originating from the  beam pipe will be parallel at
this point  and will cause a  peak in the  invariant mass distribution
around $2m_e$. Particles originating  from a decay at the origin will have an
offset in the  azimuthal angle, causing an offset in
the  invariant   mass  distribution.  This  can  be   seen  in  
Fig.~\ref{fig:conv},  where  the  conversion  events  have  a  distribution
starting at  zero while the  $\eta\rightarrow e^{+}e^{-}\gamma$ events
are offset.

The selection condition used is illustrated in Fig.~\ref{fig:conv}. As
part of a consistency check for each studied decay channel the
selection condition was varied systematically and the effect on the
final result was determined to be negligible.  This involved changing
both the slope and $y$-intercept of the diagonal component of the
demarcation line illustrated in Fig.~\ref{fig:conv}, as well as the
height of the horizontal component.

\section{Normalization: $\eta\rightarrow\pi^{+}\pi^{-}\pi^{0}_{\gamma\gamma}$}

The channel $\eta\rightarrow\pi^{+}\pi^{-}\pi^{0}$ has a branching
ratio of $0.2292 \pm 0.0028$, making it the most probable decay of the
$\eta$ meson containing charged particles in the final state
\cite{PDG}.  Due to the large branching ratio, a data sample with high
statistics and low background could be extracted. The
decay kinematics of this channel have been the subject of detailed studies
using the WASA-at-COSY 2008 data. The results have been recently
reported \cite{Adlarson:2014aks}. In the present analysis the decay
serves as a normalization channel for the less abundant processes.

\begin{figure}[!h] 
\begin{center}
 \includegraphics[width=0.45\textwidth]{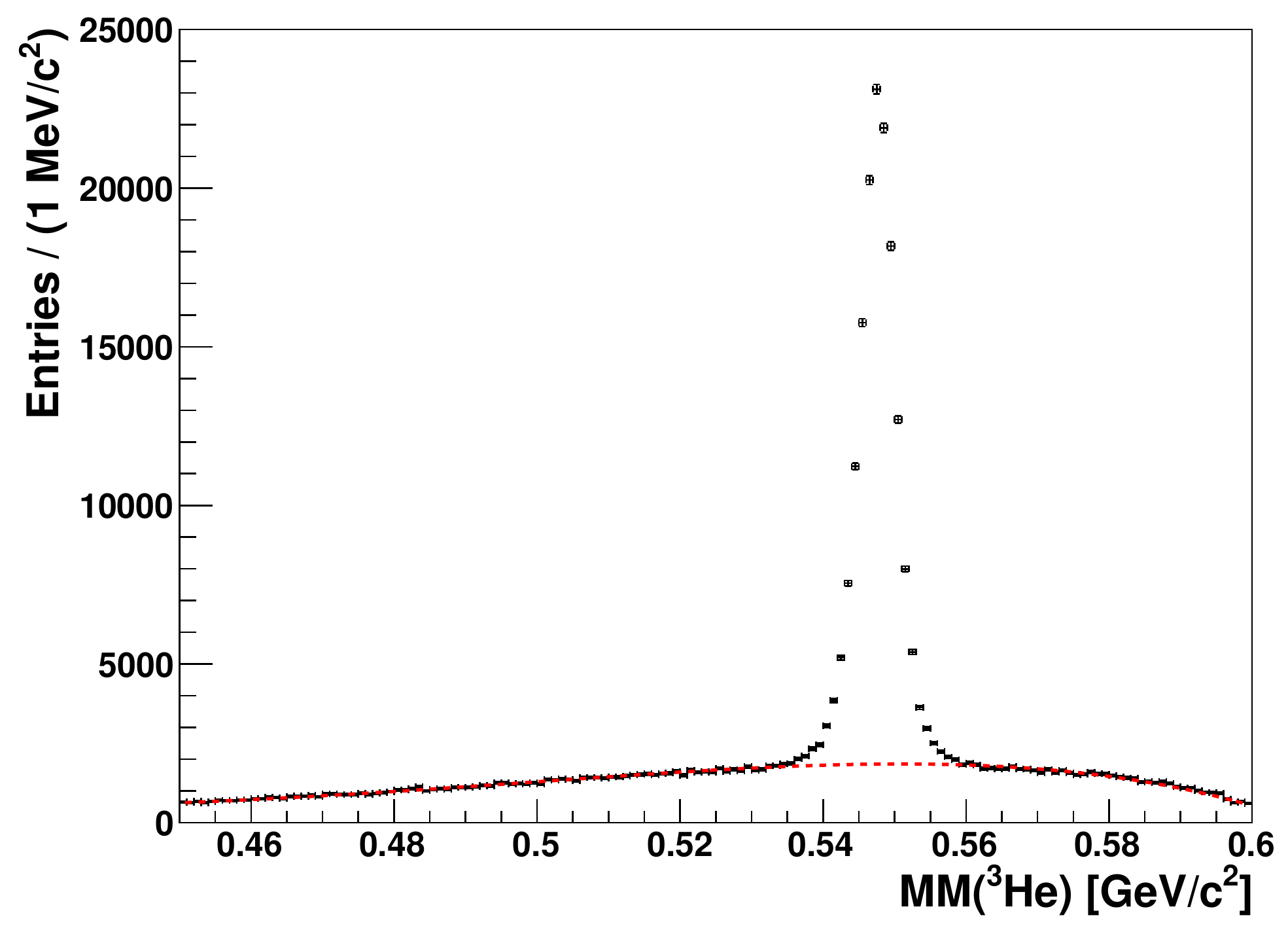}
 \caption[Selection of $\eta\rightarrow\pi^{+}\pi^{-}\pi^{0}$]{(Color online) Missing mass of ${}^{3}\textrm{He}$ for events passing the kinematic fit probability condition for the hypothesis ${pd}\rightarrow {^{3}\textrm{He}}\,\pi^{+}\pi^{-}\gamma\gamma$. The background fit is described by a dashed line and is derived using the method described in the text.}
\label{fig:3piC}
\end{center}
\end{figure}

The MM$({^{3}\textrm{He}})$ distribution for events passing the
kinematic fit condition for the
$pd\to{^{3}\textrm{He}}\,\pi^{+}\pi^{-}\gamma\gamma$ hypothesis is shown
in Fig.~\ref{fig:3piC}.  Due to the large signal to background ratio,
no additional selection conditions are required.  The $\eta$ peak after
subtraction of the continuous background is expected to contain only a
2\% contribution from $\eta\rightarrow\pi^{+}\pi^{-}\gamma$ with one
spurious neutral cluster.
\label{sec_bgfit}
The smooth background under the peak in Fig.~\ref{fig:3piC} is
composed mostly of ${pd}\rightarrow{}^{3}\textrm{He}\,\pi^{+}\pi^{-}\pi^{0}$ events
with a small contribution from ${pd}\rightarrow{}^{3}\textrm{He}\,\pi^{+}\pi^{-}$
events. In order to fit the background, 
the MM$({^{3}\textrm{He}})$ spectra of these two
processes are determined from Monte Carlo simulations assuming a
homogenous phase space distribution. They are multiplied by
a fourth-order polynomial to model the acceptance and possible deviations
from the phase space distributions. An additional
parameter controls the relative scaling of the two-
and three-pion continuum spectra. When the fit is performed the region
$\pm3\sigma$ around the peak is excluded. The number of signal events
is then determined by subtracting the background function from the
experimental spectrum in the signal area $\pm3\sigma$ around the $\eta$ peak and
integrating the resulting spectrum. Using this method $(136,240\pm
410_{stat})$ and $(290,810\pm 590_{stat})$
$\eta\rightarrow\pi^{+}\pi^{-}\pi^{0}_{\gamma\gamma}$ events are found
for the 2008 and 2009 data sets, respectively.

The same method of background subtraction is used for the other
channels, with the exception of $\eta\rightarrow e^{+}e^{-}e^{+}e^{-}$
where the selection conditions are stringent enough to reject nearly
the entire continuum background. The systematic error of the fitting procedure 
is determined by varying the method used. One method is
identical to that described above but includes the line shape of
the $\eta$ peak, determined from simulations, along with an extra scaling parameter. Third and
fourth order polynomials are also used to model the background, with both the fit range and
exclusion range systematically varied.

\section{$\eta\rightarrow e^{+}e^{-}\gamma$}
\label{sec_etaeeg}

The $\eta \to e^{+}e^{-}\gamma$ Dalitz decay proceeds via a real and
a virtual photon intermediate state with the virtual photon converting
into an $e^{+}e^{-}$ pair. According to the vector-meson dominance
model, the virtual photon can mix with neutral vector mesons.  This
mixing is dominated by the $\rho$-meson ($m=775.26\pm0.25$ MeV, $\Gamma=149.1\pm0.8$ MeV) with 
contributions from the tails of the $\omega$-meson ($m=782.65\pm0.12$ MeV, $\Gamma=8.49\pm0.08$ MeV) and $\phi$-meson ($m=1019.461\pm0.019$ MeV, $\Gamma=4.266\pm0.031$ MeV) distributions. The squared
four-momentum of the virtual photon corresponds to the
squared invariant mass of the $e^{+}e^{-}$ pair, the
invariant mass distribution of $e^{+}e^{-}$ pairs is affected by this
mixing and the transition form factor can be extracted (see for example
Refs. \cite{Berghauser2011,PhysRevC.89.044608}). In this publication
we present only results on the branching ratio while
assuming the transition form factor according to the vector meson dominance model.

The branching ratio of this  channel given by Ref. \cite{PDG} is $(6.9
\pm    0.4)\times   10^{-3}$    based   on    the    measurements   in
Refs. \cite{Akhmetshin2001,Achasov2001,Berlowski, Berghauser2011}.
The  largest data  samples to date consist of 
$(1345\pm59)$ and $(2.2\times10^{4})$ events \cite{Berghauser2011,PhysRevC.89.044608}. This channel
was  also  measured  in  the  $pd\rightarrow  {}^{3}\textrm{He}\,\eta$
reaction  by the CELSIUS/WASA  collaboration with  $(435\pm35)$ events
collected \cite{Berlowski}.

The first selection condition  is a  threshold of 100 MeV
for the  energy of  the photon. The cut does not reduce  signal  efficiency,  but  significantly  reduces  the
contribution    of   the pion   background    where
$\pi^{+}\pi^{-}$  pairs  are combined  with  a spurious neutral  cluster. A 
neutral low energy cluster  in the calorimeter can come from noise or
coincidental background that has not been rejected by the standard analysis.

After the cut the  fraction of the  events in  the  MM$({^{3}\textrm{He}})$
$\eta$  peak    is    70\%    the    signal    channel,    5\%
$\eta\rightarrow\pi^{+}\pi^{-}\gamma$,   and   24\%   
$\eta\rightarrow
\gamma\gamma$,     with    a     small    remainder     coming    from
$\eta\rightarrow\pi^{+}\pi^{-}\pi^{0}$. 
These numbers are  determined by  studies of  Monte Carlo
simulations. 

In  order  to  reduce   the  photon  conversion  background  from  the
$\eta\rightarrow\gamma\gamma$   decay,    the   conversion   condition
introduced in Section \ref{sec_conv} is applied. Simulations show that
this  reduces  the  contribution of  $\eta\rightarrow\gamma\gamma$  to
nearly zero, while reducing the number of signal events by 20\%.
The background       from
$\eta\rightarrow\pi^{+}\pi^{-}\gamma$ is rejected  by using the particle identification 
rejection algorithm 
presented  in  Section \ref{sec_pid}.   This  reduces the
background contribution  to about  1\%
with about a 10\% decrease in the number of the signal events. The
final  MM$({^{3}\textrm{He}})$ $\eta$ peak consists of over 98\%  $\eta\to e^{+}e^{-}\gamma$
events.

\begin{figure}[!h] 
\begin{center}
 \includegraphics[width=0.5\textwidth]{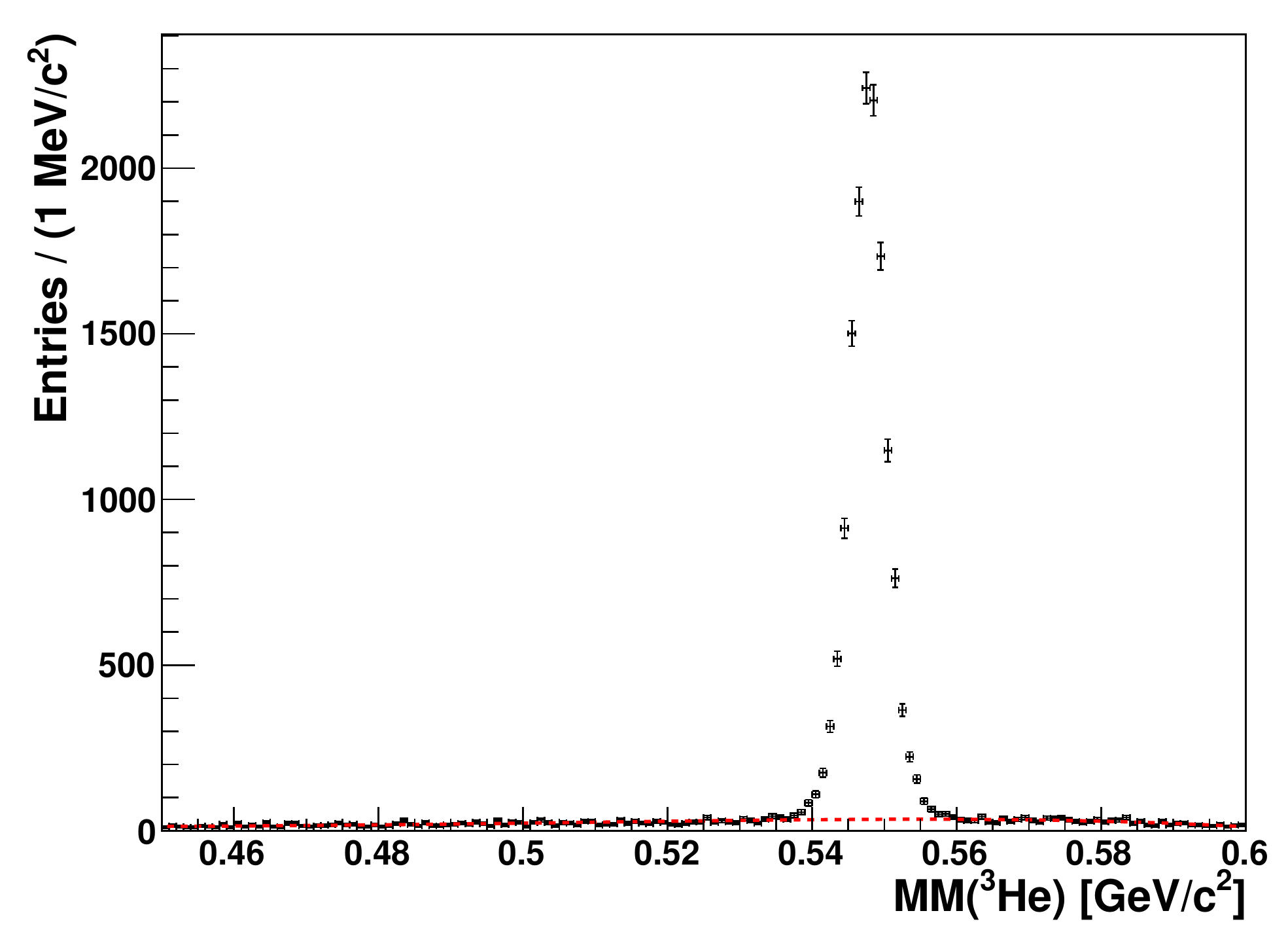}
 \caption[Event sample]{(Color online) ${^{3}\textrm{He}}$ missing mass distribution for events passing all selection conditions for $\eta\rightarrow e^{+}e^{-}\gamma$. The background fit is described by the dashed line.}
\label{fig:eventseeg}
\end{center}
\end{figure}

After all  conditions are applied  the number of events  is determined
from the $\eta$ peak content of the MM$({^{3}\textrm{He}})$ distribution 
(Fig.~\ref{fig:eventseeg}).   After subtracting  the   small  remaining
background  from   competing  $\eta$ decay channels   the  
peak contains $(14,040\pm120_{stat})$  $\eta\rightarrow e^{+}e^{-}\gamma$ events. 

\label{sec:syseeg}
The systematic error was determined by varying the selection conditions and checking if the specific choice for any condition has a systematic effect on the result.
 
\begin{compactitem}
  \item[\it{$e^{+}e^{-}$ identification}:] The default condition is a probability of at least 50\% that the particles are $e^{+}e^{-}$. This is varied from 30\% to 70\%.
  \item [\it{Photon conversion}:] Both the slope of the diagonal line and the height of the horizontal line composing the selection demarcation (see Fig. \ref{fig:conv}) are varied systematically. 
  \item [\it{Kinematic fit probability}:] The condition on the probability of the kinematic fit is varied from its default of 10\% up to 35\% for the signal and normalization channels simultaneously. Conditions below 10\% are not useful because the probability distribution is not flat in this region and includes a large amount of background, making extraction of the signal difficult. 
  \item [\it{Instantaneous luminosity}:] The instantaneous luminosity is monitored by recording the rates of the elastic scattering trigger and the pellet target every second.  The branching ratio is extracted in bins of the elastic scattering trigger normalized to the target rate. The normalization to the target rate is needed to account for the duty factor of the pellet target.
  \item [{\it Continuous background subtraction}:] Several methods of background subtraction are applied as  described in Section \ref{sec_bgfit}. 
\item [{\it Calibration and luminosity}:] The data sets from 2008 and 2009 differ only by calibration and luminosity profile.
These parameters are included in the detector simulation and their systematical uncertainty is estimated
by separate analysis of the two data sets. This leads to assignment of 4\% systematic uncertainty due to deviations between the two data sets.
\end{compactitem}

The methodology from Ref. \cite{Barlow} was applied to compare the significance
of a proposed systematic effect to the differences in statisical error of the subsets of data used to 
derive this significance. Using this
method on the the first four conditions above, none exhibit a significance exceeding
3$\sigma$ and a systematic error is not assigned. The final test, differences in calibration 
and luminosity profile between the 2008 and 2009 datasets, shows a 4\% deviation common to all channels. This error has been included in the final results.

It is not possible to check the effect of the background subtraction
using this method. Therefore, the systematic error on the background
subtraction was determined by performing several different fits of the 
background and taking the standard deviation.  The error on the acceptance due to the
uncertainties on the transition form factor is determined to be negligible by performing the 
analysis using simulated data with several different values for the transition form factor.

\label{sec:breeg}

The branching ratio  relative to $\eta\rightarrow\pi^{+}\pi^{-}\pi^{0}$ is obtained after correcting for the respective backgrounds and the final acceptance of 12.0\% for the signal channel:
\begin{equation*}
 \begin{split}
 \Gamma(\eta\rightarrow e^{+}e^{-}&\gamma)/\Gamma(\eta\rightarrow\pi^{+}\pi^{-}\pi^{0}_{\gamma\gamma}) =\\ 
 &  (2.97\pm0.03_{stat/fit}\pm0.13_{sys})\times 10^{-2}
\end{split}
\end{equation*}

The resulting branching ratio is in reasonable agreement with other experimental values \cite{PDG}, with 
a precision limited by the systematic error resulting from the luminosity profile.

\section{$\eta\rightarrow e^{+}e^{-}e^{+}e^{-}$}
\label{sec_etaeeee}
The decay $\eta\rightarrow e^{+}e^{-}e^{+}e^{-}$ is closely related to the decay 
$\eta\rightarrow e^{+}e^{-}\gamma$ above and proceeds via two virtual photons. 
The additional electromagnetic coupling suppresses the branching ratio of the decay
by two orders of magnitude compared to $\eta\rightarrow e^{+}e^{-}\gamma$. The only
measurement where this decay is observed was performed by the KLOE Collaboration. 
The branching ratio was determined to be $(2.4\pm0.2\pm0.1)\times10^{-5}$ 
based on $(362 \pm 29)$ events \cite{Ambrosino2011a}. 

Event candidates with at least two positively and two negatively
charged particles measured in the WASA central detector are passed through
a kinematic fitting routine with the $pd\rightarrow
{^{3}\textrm{He}}\,e^{+}e^{-}e^{+}e^{-}$ hypothesis. Only events
fulfilling energy and momentum conservation at greater than 10\%
probability are further considered.

To suppress the charged-particle background from
$\pi^{\pm}$-mesons the particle identification rejection algorithm
introduced in Section \ref{sec_pid} is used. Since the algorithm
considers one positively and one negatively charged particle pair at
once, an $e^{+}e^{-}e^{+}e^{-}$ event candidate is accepted if both
pairs in the two possible combinations of oppositely charged
particles passes the selection condition.

Background from photon conversion in the beam pipe, predominantly from
the reaction $\eta\rightarrow e^{+}e^{-}\gamma$, is suppressed using
the method presented in Section \ref{sec_conv}. Again an event is
accepted if both pairs in  the two possible combinations pass the
condition. 

\begin{figure}[!h]
\begin{center}
 \includegraphics[width=0.5\textwidth]{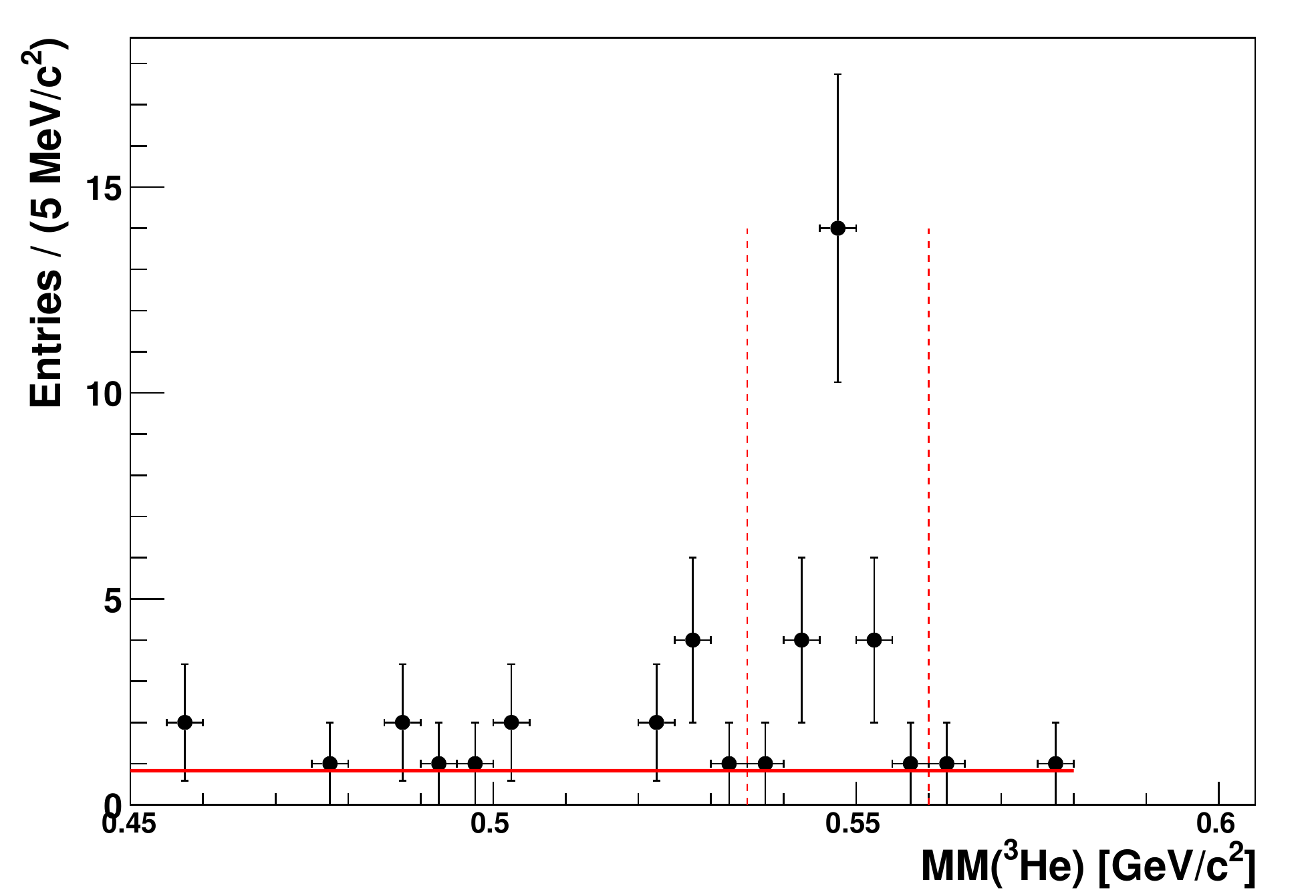}
 \caption[Final Event Sample eeee]{ (Color online) The MM$({^{3}\textrm{He}})$
   distribution for events passing all selection conditions for
   the $\eta\to e^{+}e^{-}e^{+}e^{-}$ signal. The background coming from
   $\pi$-meson production is described by the horizontal red line. The region
   between the vertical dashed lines was excluded from the fit of the background.}
\label{fig:eeee}
\end{center}
\end{figure}

After applying these criteria, the remaining number of $\eta$ events
is extracted from the MM$({^{3}\textrm{He}})$ spectrum shown in 
Fig.~\ref{fig:eeee}. Due to the limited statistics a
simplified method with constant continuous background term is used in
the fit. The range from 0.535 to 0.560 GeV/$c^{2}$ is excluded from
the fit (marked by the two dashed lines in the distribution). The
number of remaining $\eta$ events is $19.7\pm4.9_{stat}$, which is
determined by counting the events in the signal region after
subtracting the background fit.

Background from other $\eta$ meson decays is determined from Monte
Carlo simulations. The only channel found to contribute
with at least one event is $\eta\rightarrow e^{+}e^{-}\gamma$. After
subtraction of the $\eta$-decay background,
$18.4\pm4.9_{stat}$ events remain.

Each selection condition was studied to
identify possible systematic effects on the branching ratio.  The
checks for systematic effects include $e^{+}e^{-}$ identification,
photon conversion and kinematic fit probability as
described in Section \ref{sec:syseeg}.  None of these checks produces an effect
with a significance exceeding
3$\sigma$, so a systematic error is not assigned. The systematic error included in the result 
comes from the 4\% error assigned due to differences 
in the calibration and luminosity profiles of the two data taking periods, as well as the error determined by using different background fits.

The branching ratio relative to $\eta\rightarrow\pi^{+}\pi^{-}\pi^{0}$
is obtained after correcting for the respective backgrounds and the final 3.3\% acceptance for the 
signal channel:

\begin{equation*}
 \begin{split}
 \Gamma(\eta\rightarrow e^{+}e^{-}&e^{+}e^{-})/\Gamma(\eta\rightarrow\pi^{+}\pi^{-}\pi^{0}_{\gamma\gamma}) =\\ 
 & (1.4 \pm 0.4_{stat} \pm 0.2_{sys}) \times 10^{-4}
\end{split}
\end{equation*}

This is only the second analysis of this channel to reach a finite value of the branching ratio. 
The result is compatible within errors to the previous analysis \cite{Ambrosino2011a}. 

\section{$\eta\rightarrow\pi^{+}\pi^{-}\gamma$}
\label{sec_etapipig}


The decays $\eta\rightarrow\pi^{+}\pi^{-}\gamma$ and
$\eta\rightarrow\pi^{+}\pi^{-}e^{+}e^{-}$ are driven by the same
underlying mechanism, corresponding to anomalous terms in the QCD
action. These anomalies are described by the Wess-Zumino-Witten
Lagrangian, which contains two terms pertinent for the $\eta$ decays
\cite{Wess,Witten1983422}. The so-called ``triangle'' and ``box''
anomalies describe respectively the coupling of a pseudoscalar to two
vectors and the coupling of a pseudoscalar to two pseudoscalars and
one vector. The names are inspired by the shapes of the corresponding
Feynman diagrams. The $\eta\rightarrow\pi^{+}\pi^{-}\gamma^{(*)}$
reaction is described at the lowest order of the chiral perturbation
theory entirely by the box-anomaly. However, within the framework of
the vector-meson dominance model, the triangle anomaly will dominantly 
contribute since the $\pi^{+}\pi^{-}$ pair in $P$-wave comes from
the $\rho^0$ meson contribution.

Various theoretical approaches attempt to determine the relative
contributions from these diagrams and in particular to predict the
contribution of the box diagram for the two observables: the branching
ratio and the shape of the $\pi^{+}\pi^{-}$ invariant mass spectrum
\cite{Petri2010,Bijnens1990488,Picciotto1992,Benayoun2003525,
  Venugopal19984397,Holstein200255,Borasoy2004362, Stollenwerk:2011zz}.  The channel is
the second most probable $\eta$ decay channel to charged particles with a branching ratio of
$(4.22\pm0.08)\times10^{-2}$ \cite{PDG}. It was studied by few
experiments \cite{Thaler1973, Gormley1970,Lopez:2007ab,
  WASA_pipig,Ambrosino2011}. The two most recent results: from
WASA-at-COSY (using the 2008 $pd$ data) \cite{WASA_pipig}
and from KLOE \cite{Ambrosino2011}, provide the $\pi^{+}\pi^{-}$
invariant mass spectrum with sufficient precision to see an influence of
the box diagram contribution.  The branching ratios normalized to the
$\eta\rightarrow\pi^{+}\pi^{-}\pi^{0}$ decay from CLEO
\cite{Lopez:2007ab} and KLOE \cite{Ambrosino2011} collaborations are
significantly below previous values. 

\begin{figure}[!h]
\begin{center}
 \includegraphics[width=0.45\textwidth]{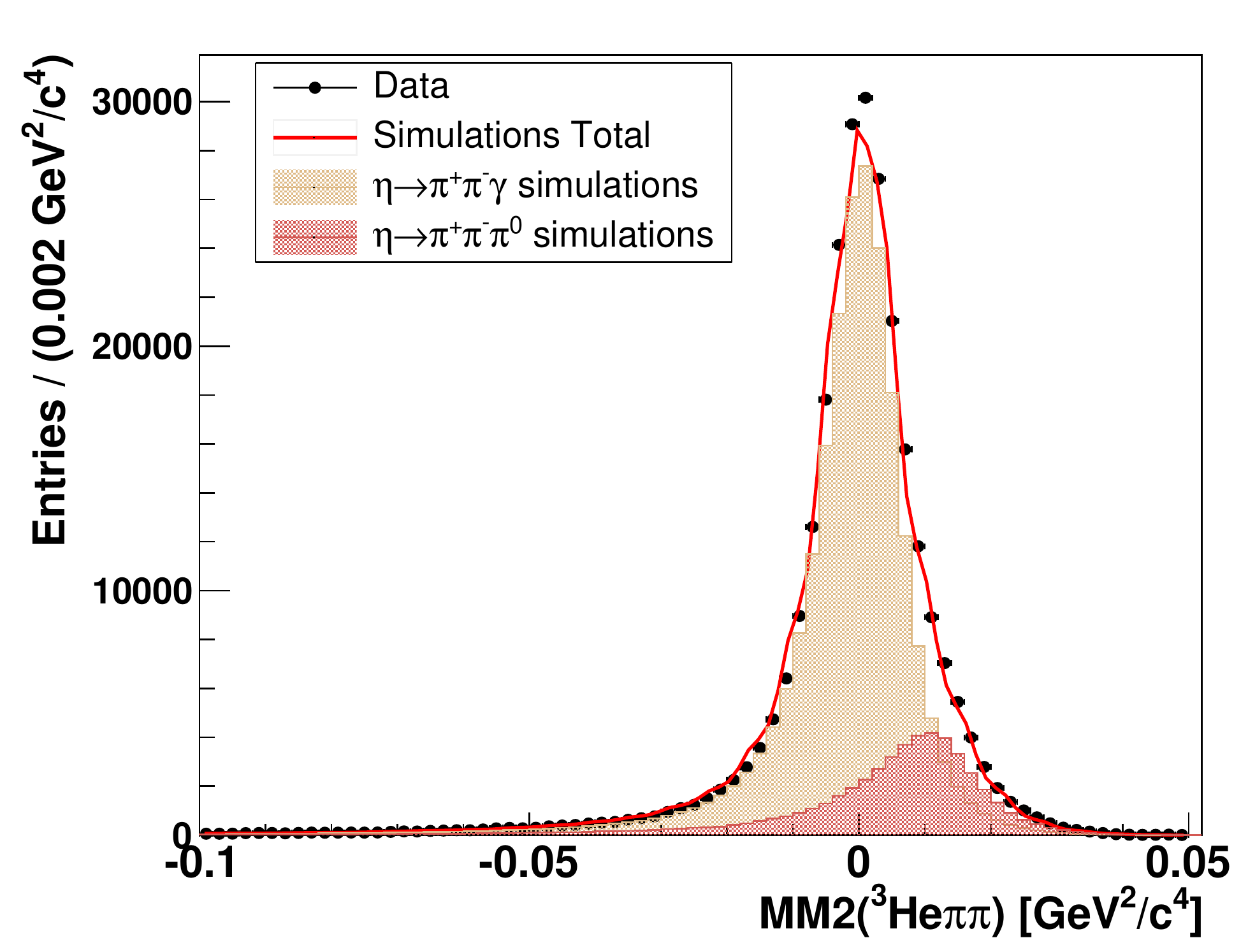}
 \caption[eta content]{(Color online) The experimental MM$^{2}({^{3}\textrm{He}}\,\pi^{+}\pi^{-})$
distribution for the $pd\rightarrow {^3\textrm{He}}\,\eta$ events after the kinematic fit probability
cut for the $pd\rightarrow {^3\textrm{He}}\,\pi^{+}\pi^{-}\gamma$ hypothesis (points)
compared to MC of the $\eta\to\pi^{+}\pi^{-}\gamma$ (light shaded) signal and the $\eta\to\pi^{+}\pi^{-}\pi^0$
background (dark shaded). The red curve is the sum of the simulations. Kinematic variables used to obtain the distributions are not corrected by the kinematic
fit.}
\label{fig:MM23Hepp}
\end{center}
\end{figure}

Events are selected with at least two oppositely-charged particles and one neutral
particle fulfilling the kinematic fit requirement for the $pd\rightarrow
{^{3}\textrm{He}}\,\pi^{+}\pi^{-}\gamma$ hypothesis.  At this point the
content of the $\eta$ peak in the MM$({^{3}\textrm{He}})$ spectrum is
composed of 70\% $\eta\rightarrow\pi^{+}\pi^{-}\gamma$, with the
remaining background mostly due to
$\eta\rightarrow\pi^{+}\pi^{-}\pi^{0}$ events where one photon is not
detected. This contribution can be reduced by placing a condition on
the missing mass squared of ${^{3}\textrm{He}}$, $\pi^{+}$, and
$\pi^{-}$, MM$^{2}({^{3}\textrm{He}}\,\pi^{+}\pi^{-})$. For the signal
channel $\eta\rightarrow\pi^{+}\pi^{-}\gamma$ the
MM$^{2}({^{3}\textrm{He}}\,\pi^{+}\pi^{-})$ distribution peaks at zero,
while for $\eta\rightarrow\pi^{+}\pi^{-}\pi^{0}$ it peaks at the
squared mass of the $\pi^{0}$. Rejection of the events with
MM$^{2}({^{3}\textrm{He}}\,\pi^{+}\pi^{-})>0.005$ GeV$^2$/c$^4$ increases
the signal content of the $\eta$ peak to 91\%.
The effect of the cut and the impact of the remaining $\eta\to\pi^{+}\pi^{-}\pi^{0}$ contribution 
is illustrated in 
Fig.~\ref{fig:MM23Hepp}. The experimental points correspond to the $\eta$
peak content determined from the MM$({^{3}\textrm{He}})$ distributions for each  
MM$^{2}({^{3}\textrm{He}}\,\pi^{+}\pi^{-})$ bin.

The non-resonant background comes predominantly from the
$pd\rightarrow {^3\textrm{He}}\,\pi^{+}\pi^{-}$ reaction where a
spurious photon is detected.  Reduction of this contribution decreases
both systematic error on the background fit and the statistical error
of the final result. A major source of the spurious photons comes from
so-called hadronic splitoffs. This happens when an interaction or a decay of 
one of the charged $\pi$-mesons creates a secondary particle which
leaves a signal in an isolated calorimeter module. In this case a spurious neutral cluster is reconstructed.

A condition to reduce the contribution of the splitoffs is applied for
the photon candidates with low energy which are close to the expected
impact point of the charged pion track in the calorimeter.  The
condition was optimized to minimize the statistical error of the extracted
number of the signal events.

\begin{figure}[!h]
\begin{center}
 \includegraphics[width=0.45\textwidth]{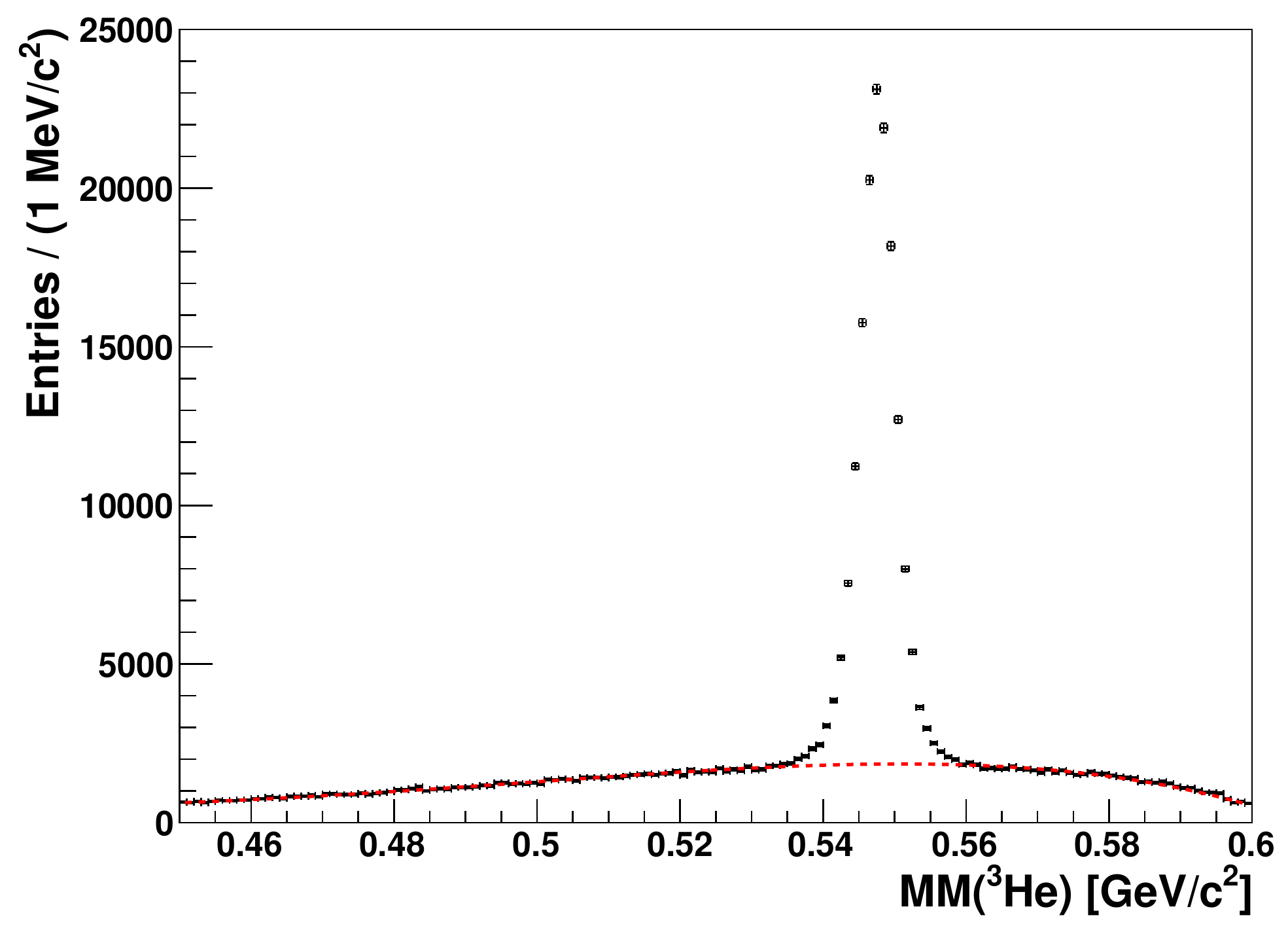}
 \caption[Final Event Sample]{(Color online) The MM$({^{3}\textrm{He}})$ distribution for
   events passing selection conditions for
   $\eta\rightarrow\pi^{+}\pi^{-}\gamma$. The function used to
   subtract the continuum background is shown as a dashed line.}
\label{fig:eventspipig}
\end{center}
\end{figure}

The MM$({^{3}\textrm{He}})$ distribution for the final selection is
shown in Fig.~\ref{fig:eventspipig}.  The background fit is
performed using the methods described in Section \ref{sec_bgfit} and
after subtraction of the $\eta\rightarrow\pi^{+}\pi^{-}\pi^{0}$
contribution to the $\eta$ peak the number of signal events is
$(139,760 \pm 430)$.

No variation exceeding 3$\sigma$ for the checks due to kinematic fit
probability, instantaneous luminosity, described in Section
\ref{sec:syseeg}, is observed.  The calibration and luminosity
comparison of the 2008 and 2009 data sets leads to assignment of 4\%
systematic uncertainty as in the $\eta\rightarrow e^{+}e^{-}\gamma$
analysis (see Section \ref{sec:syseeg}).  The continuous background
subtraction is also investigated as for $\eta\rightarrow
e^{+}e^{-}\gamma$, with both polynomials and the methods from Section
\ref{sec_bgfit}.

The two specific conditions are investigated separately:
\begin{compactitem}
  \item[\it{Missing mass squared cut}:] The cut was varied in steps in the region between the $\pi^{0}$ mass squared and the signal peak at zero. 
  \item [\it{Splitoffs}:] The selection condition used to reject splitoffs was removed from the analysis chain and the result remains consistent. 
\end{compactitem}
The above two tests show that the conditions do not introduce
systematic deviations and therefore overall systematic error is 
determined by the background subtraction and the difference between
the two data sets.

The branching ratio normalized to
$\eta\rightarrow\pi^{+}\pi^{-}\pi^{0}_{\gamma\gamma}$:
\begin{equation*}
 \begin{split}
 \Gamma(\eta\rightarrow \pi^{+}\pi^{-}&\gamma)/\Gamma(\eta\rightarrow\pi^{+}\pi^{-}\pi^{0}_{\gamma\gamma}) =\\ 
 & 0.206\pm0.003_{stat/fit}\pm0.008_{sys}
\end{split}
\end{equation*}

\section{$\eta\rightarrow \pi^{+}\pi^{-}e^{+}e^{-}$}
\label{sec_etappee}
The decay $\eta\rightarrow\pi^{+}\pi^{-}e^{+}e^{-}$ is closely related
to $\eta\rightarrow\pi^{+}\pi^{-}\gamma$ and corresponds to the
conversion of the virtual photon leading to about a factor of ${\alpha}$
suppression. Therefore the measurement of this branching ratio provides additional
information on the mechanism contributing to the parent process, 
$\eta\rightarrow\pi^{+}\pi^{-}\gamma^{*}$. However, the small decay
probability ($O (10^{-4})$) has made the channel difficult to detect
until recently. The process has been observed by several experiments
\cite{Bargholtz2007, Ambrosino2009, Akhmetshin2001, Grossman1966}, but
the only measurement of the branching ratio with statistical
significance more than 3$\sigma$ is a recent result from the KLOE
collaboration with 1555$\pm 52$ events leading to a branching ratio of $(2.68
\pm 0.09_{stat} \pm 0.07_{sys}) \times 10^{-4}$ \cite{Ambrosino2009}.

\begin{figure}[!h]
\begin{center}
 \includegraphics[width=0.25\textwidth]{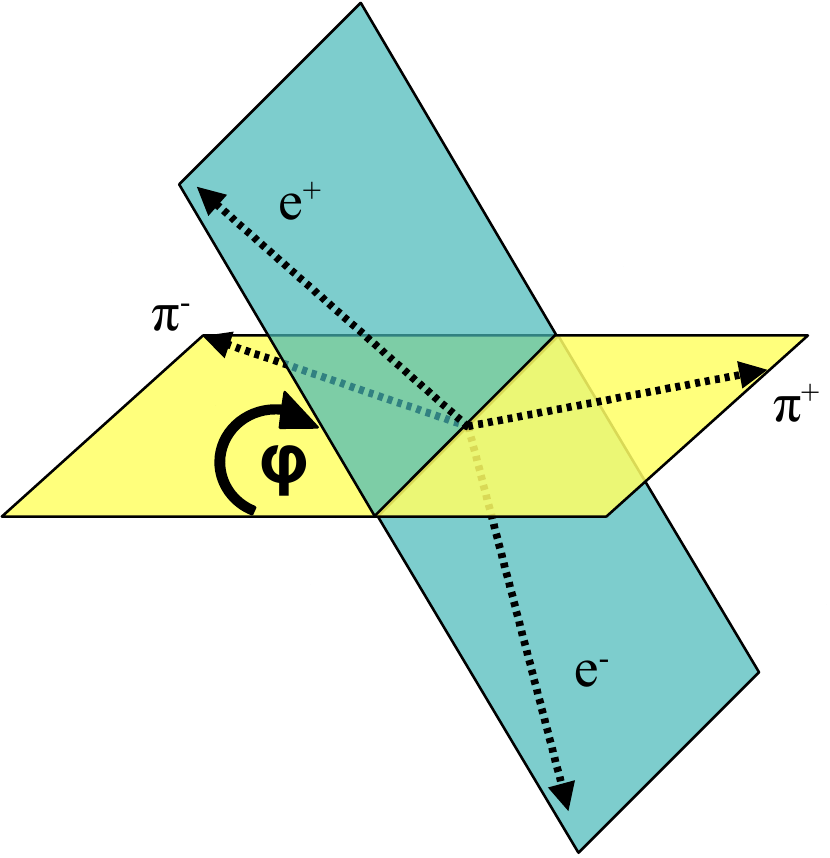}
 \caption[Dihedral Angle]{(Color online) Definition of the dihedral
   angle $\phi$ for the $\eta\rightarrow\pi^{+}\pi^{-}e^{+}e^{-}$
   decay in the $\eta$ meson rest frame. }
\label{fig:dihedral}
\end{center}
\end{figure}

The channel is also interesting due to searches for a possible
$CP$-violation mechanism outside of the Standard Model
\cite{Gao,Geng}. It has been shown that a contribution to the decay
amplitude from the $CP$-violating electric transition would result in a
linear polarization to the virtual photon. A non-zero polarization of
the virtual photon contributes to an asymmetry of the distribution of
the angle, $\phi$ (the dihedral angle), between the electron and
$\pi$-meson decay planes in the $\eta$ meson rest frame
\cite{Gao}. The $\phi$ angle is shown in Fig.~\ref{fig:dihedral}.
The asymmetry, $A_{\phi}$, is defined as:

\begin{equation*}
 A_{\phi} = \frac{N(\sin\phi\cos\phi > 0)-N(\sin\phi\cos\phi < 0)}{N(\sin\phi\cos\phi > 0)+N(\sin\phi\cos\phi < 0)}
\end{equation*}	
where $N(...)$ is the number of the decays fulfilling the corresponding
condition.

The theoretical upper limit for $A_{\phi}$ is determined by
constraints on the strong $CP$-violation from neutron electric dipole
moment measurements to be about 1\% \cite{Gao}.  A previous
measurement from the KLOE collaboration of $A_{\phi} = (-0.6 \pm 2.5
_{stat} \pm 1.8_{sys})\times 10^{-2}$ constrains the asymmetry
$|A_{\phi}|$ to be less than a few percent \cite{Ambrosino2009}.

The analysis follows the steps outlined in Section \ref{3_Analysis}
using event candidates with at least two positive and two negative
reconstructed tracks in the MDC.  A unique aspect of this decay
channel is that the final state contains both charged
$\pi$-mesons and leptons. The kinematic fitting assumes the
${pd}\rightarrow {^{3}\textrm{He}}\,\pi^{+}\pi^{-} e^{+}e^{-}$
hypothesis and all four possible mass assignments are tested. The
events with probability above 0.1 for at least one of the combinations
are accepted for the further analysis.

All four combinations for the selected events are evaluated according
to the particle identification routine described in Section
\ref{sec_pid}. Additional information about decay angles between the
oppositely-charged pairs is included in the algorithm. The angle
between the leptons is expected to be small compared to the angle
between the $\pi$-mesons. This feature was previously used in Ref.
\cite{Bargholtz2007}. The simulations of the decay $\eta\to
\pi^{+}\pi^{-} e^{+}e^{-}$ with the matrix element from
Ref. \cite{Petri2010} are used to determine the probabilities for the
correct identification of the $e^{+}e^{-}$, $\pi^{+}\pi^{-}$, and
$\pi^{\pm}e^{\mp}$ pairs as a function of the opening angle. The
angular information is added to the probabilities, again using Bayes'
equation, and the configuration with the highest probability is
accepted. This method has been tested with simulations and the correct
configuration is selected in over 90\% of events.

A significant background comes from photon conversion
in the reactions $\eta\rightarrow\pi^{+}\pi^{-}\gamma$ and
$\eta\rightarrow\pi^{+}\pi^{-}\pi^{0}$.  The conversion suppression
introduced in Section \ref{sec_etaeeg} reduces the contribution of
these channels to 5\% of the $\eta$ peak in the
MM$({^{3}\textrm{He}})$ distribution.  The largest remaining
background is from the $\eta\to\pi^{+}\pi^{-}[\pi^{0}\to
  e^{+}e^{-}\gamma]$ decay chain, and constitutes 15\% of the peak. 

\begin{figure}[!h]
\begin{center}
 \includegraphics[width=0.45\textwidth]{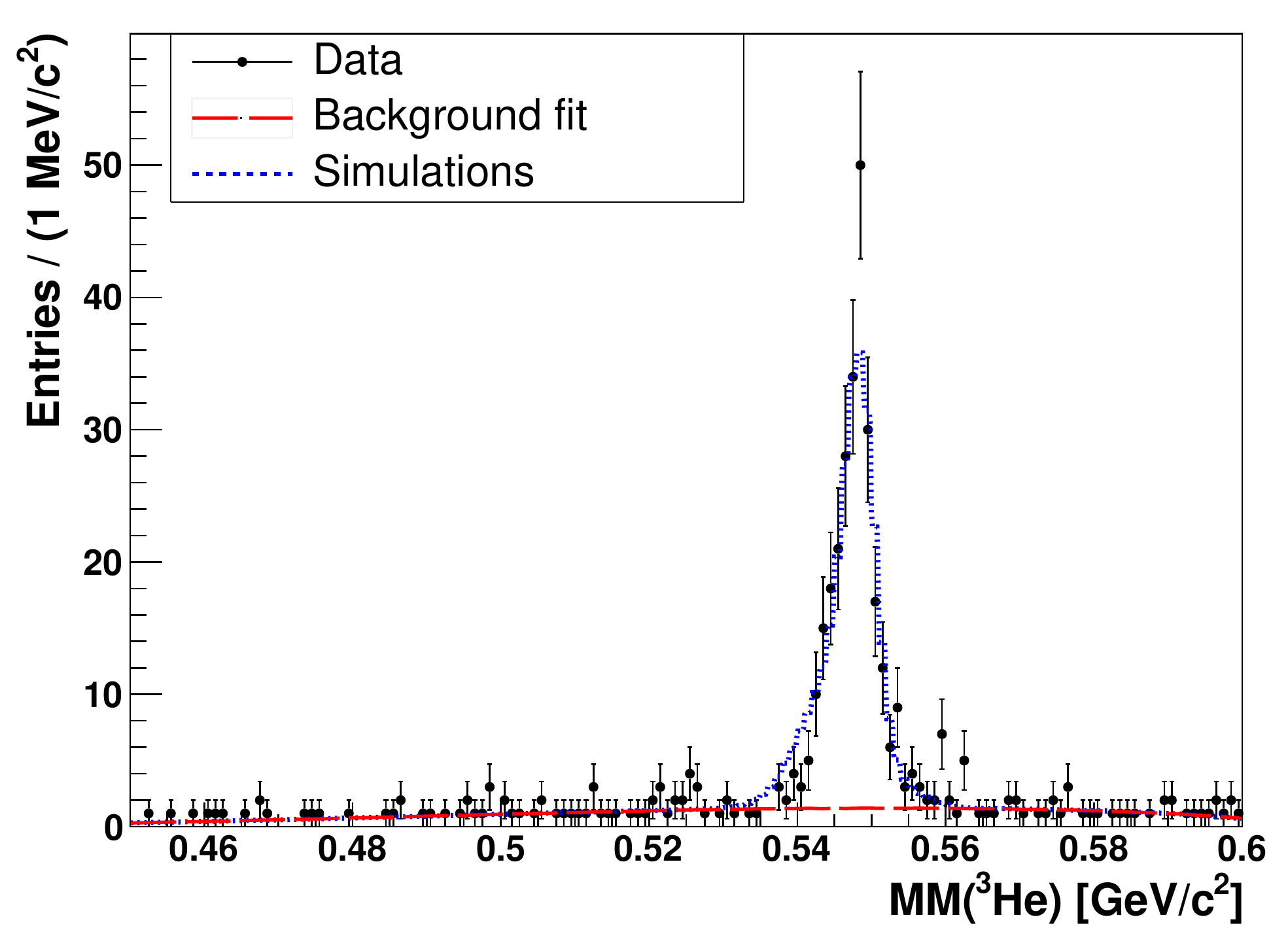}
 \caption[Final Events]{(Color online) The MM$({^{3}\textrm{He}})$
   distribution for events passing the
   $\eta\rightarrow\pi^{+}\pi^{-}e^{+}e^{-}$ selection criteria. The
   background fit is shown as a dashed line (red) and the shape of the
   peak from simulations of the ${pd}\rightarrow
   {^{3}\textrm{He}}\,\eta$ events is shown as a dotted line (blue).}
\label{fig:eventsppee}
\end{center}
\end{figure}

The missing mass distribution for events passing all selection
conditions is shown in Fig.~\ref{fig:eventsppee}.  After subtraction
of all background channels there are $(251\pm17)$ signal events found in the
combined data set.

Due to the high statistical error all of the  systematic effects from the kinematic fit probability and photon conversion
selection conditions were determined to be negligible. The 4\% error due to differences in the 2008 and 2009 data periods was 
nevertheless included as determined from the higher statistics decays. 

The systematic error on the final result is the same magnitude as the statistical error.

\begin{equation*}
 \begin{split}
 \Gamma(\eta\rightarrow \pi^{+}\pi^{-}&e^{+}e^{-})/\Gamma(\eta\rightarrow\pi^{+}\pi^{-}\pi^{0}_{\gamma\gamma}) =\\ 
 & (1.2\pm0.1_{stat}\pm0.1_{sys})\times 10^{-3}.
\end{split}
\end{equation*}

\begin{figure}[!h]
\begin{center}
 \includegraphics[width=0.45\textwidth]{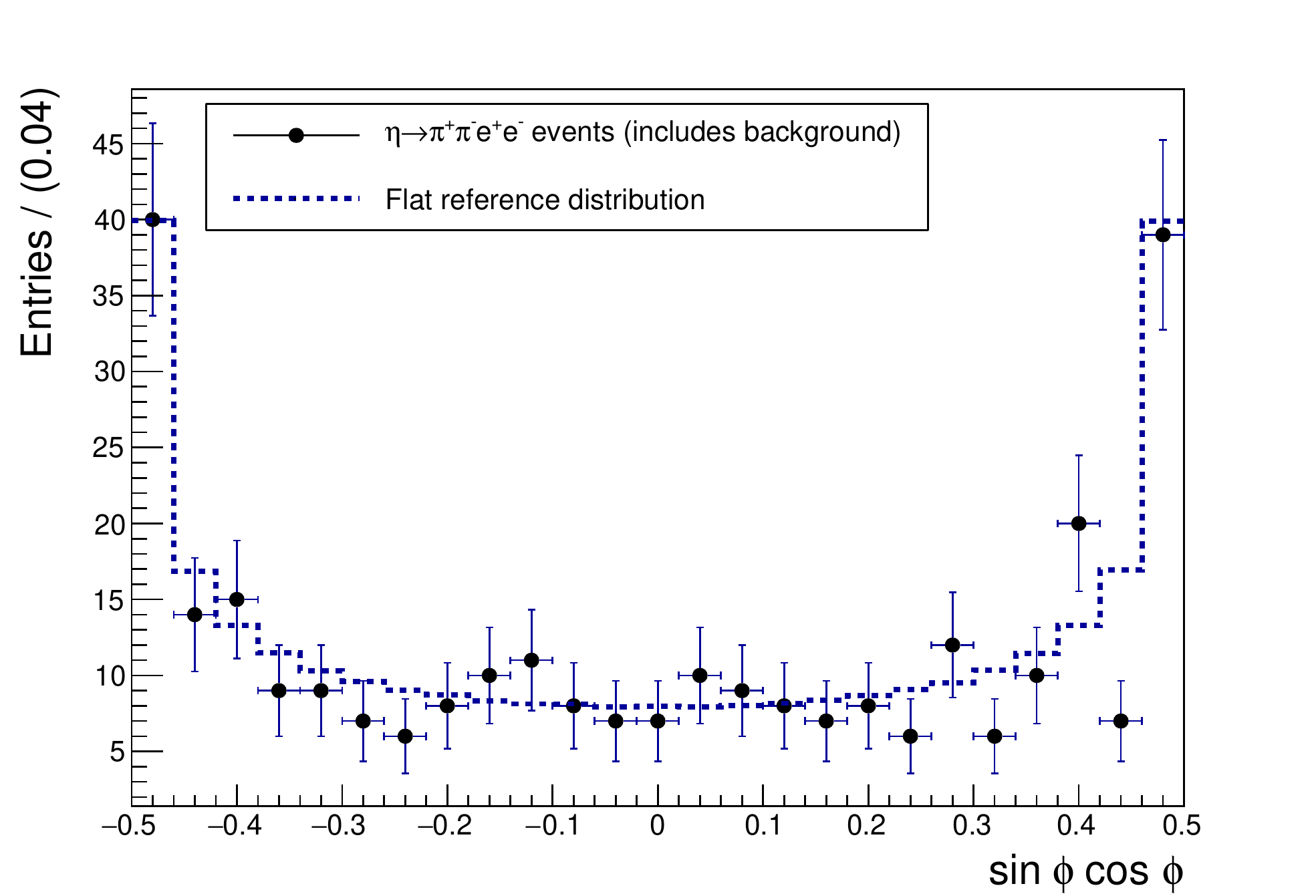}
 \caption[Dihedral Angle]{(Color online)  $\sin\phi\cos\phi$  distribution for the data and Monte Carlo simulation 
of the $\eta\rightarrow\pi^{+}\pi^{-}e^{+}e^{-}$ with a flat $\phi$ distribution. }
\label{fig:ppeeaphi}
\end{center}
\end{figure}

The angle between the $e^{+}e^{-}$ and $\pi^{+}\pi^{-}$ decay planes
was determined for each event in the final event sample using a method
presented from Ref. \cite{Ambrosino2009}. The 
$\sin\phi\cos\phi$ distribution for the selected data sample is shown in Fig.~\ref{fig:ppeeaphi}
and compared to a Monte Carlo simulation assuming a flat $\phi$ distribution. 
The data are divided into $\sin\phi\cos\phi>0$ and
$\sin\phi\cos\phi<0$ sub-samples leading to the two
MM$({^{3}\textrm{He}})$ distributions.  Due to the low magnitude of the
continuous background, the fit of the multi-pion background distribution uses a variety third and fourth order
polynomials where the fit range and peak exclusion range are changed
systematically. The $\eta$ peak content is obtained as the average
value for the fits.  The number of signal events in each class is
obtained by integrating the peak after background subtraction, further subtracting background from 
other $\eta$ decay channels determined from simulations, and
correcting the result for acceptance.

The same sources of systematic error were considered as for the
branching ratio analysis.  Only the error on the continuous background
fit is included in the final results and it is determined by the
standard deviation of the various fits. All other errors were insignificant compared to the 
statistical error.

The final result for the asymmetry: 
\begin{equation*}
 A_{\phi} = (-1.1 \pm 6.6_{stat} \pm 0.2_{sys}) \times 10^{-2}.
 \end{equation*}

\section{Conclusion}
\label{sec_conclusion}

\begin{table}[h]
\begin{center}
\begin{tabular}{ l l }

\hline
Channel  &  Branching Ratio\\
          &  w.r.t. $\eta\rightarrow\pi^{+}\pi^{-}\pi^{0}_{\gamma\gamma}$\\
\hline
$\eta\rightarrow\pi^{+}\pi^{-}\gamma$ & $0.206\pm0.003_{stat/fit}\pm0.008_{sys}$ \\
$\eta\rightarrow e^{+}e^{-}\gamma$ & $(2.97\pm0.03_{stat/fit}\pm0.13_{sys})\times 10^{-2}$ \\
$\eta\rightarrow \pi^{+}\pi^{-}e^{+}e^{-}$ & $(1.2\pm0.1_{stat}\pm0.1_{sys})\times 10^{-3}$\\
$\eta\rightarrow e^{+}e^{-}e^{+}e^{-}$ & $(1.4 \pm 0.4_{stat} \pm 0.2_{sys}) \times 10^{-4}$ \\
\hline
\end{tabular}
\caption[Absolute Branching Ratios]{Summary of experimental results
  for branching ratios relative to the normalization channel
  $\eta\rightarrow\pi^{+}\pi^{-}\pi^{0}_{\gamma\gamma}$.}
\label{t_BRsRel}
\end{center}
\end{table}

The obtained results on the relative branching ratios relative to the
normalization channel
$\eta\rightarrow\pi^{+}\pi^{-}\pi^{0}_{\gamma\gamma}$ are summarized
in Table~\ref{t_BRsRel}. The deduced value for
$\Gamma(\eta\rightarrow\pi^{+}\pi^{-}\gamma)/\Gamma(\eta\rightarrow\pi^{+}\pi^{-}\pi^{0})$
is $0.206\pm0.003_{stat/fit}\pm0.008_{sys}$.  It is in good
agreement with the older experiments \cite{Gormley1970, Thaler1973}
but is 2.6 and 2.5 standard deviations above the recent values from
CLEO \cite{Lopez:2007ab} and KLOE \cite{Ambrosino2011} respectively.

The measured relative branching ratios can be translated to
absolute branching ratios by using known world averages from
Ref.~\cite{PDG} for the branching ratios of $\eta\rightarrow
\pi^{+}\pi^{-}\pi^{0}$ and $\pi^{0}\rightarrow\gamma\gamma$.  The
results are presented in Table~\ref{t_BRsAbs}.  

The branching ratio for $\eta\rightarrow e^{+}e^{-}\gamma$ is
consistent with the most recent Particle Data Group fit $(6.9 \pm
0.4)\times10^{-3}$ but it is more precise by 20\%.  The absolute
branching ratios for $\eta\rightarrow\pi^{+}\pi^{-}e^{+}e^{-}$ and
$\eta\rightarrow e^{+}e^{-}e^{+}e^{-}$ decays are in good agreement
with the values reported by KLOE. \cite{Ambrosino2009,Ambrosino2011a}

The measured dihedral angle asymmetry, $A_\phi$ for
$\eta\rightarrow\pi^{+}\pi^{-}e^{+}e^{-}$ has been determined to be
consistent with zero: $A_\phi=(-1.1 \pm 6.6_{stat} \pm 0.2_{sys})
\times 10^{-2}$.

\begin{table}[h] 
\begin{center}
\begin{tabular}{ l r }
\hline
Channel  &  Branching Ratio\\
\hline
$\eta\rightarrow\pi^{+}\pi^{-}\gamma$ & $(4.67\pm 0.07_{stat/fit} \pm 0.19_{sys})\times 10^{-2}$ \\
$\eta\rightarrow e^{+}e^{-}\gamma$ & $(6.72 \pm 0.07_{stat/fit} \pm 0.31_{sys}) \times 10^{-3}$ \\
$\eta\rightarrow \pi^{+}\pi^{-}e^{+}e^{-}$ & $(2.7 \pm 0.2_{stat}\pm 0.2_{sys}) \times 10^{-4} $\\
$\eta\rightarrow e^{+}e^{-}e^{+}e^{-}$ & $(3.2\pm0.9_{stat}\pm0.5_{sys})\times 10^{-5}$ \\
\hline
\end{tabular}
\caption[Absolute Branching Ratios]{Summary of experimental results
  for the absolute branching ratios, extrapolated from the relative
  branching ratio for each channel with respect to
  $\eta\rightarrow\pi^{+}\pi^{-} [ \pi^{0}\rightarrow\gamma\gamma]$
  using the branching ratios from Ref. \cite{PDG}:
  $BR(\eta\rightarrow\pi^{+}\pi^{-}\pi^{0})=(2.292 \pm 0.028) \times 10^{-1}$
  and $BR(\pi^{0}\rightarrow\gamma\gamma)=(98.823 \pm 0.034) \times
  10^{-2}$.}
\label{t_BRsAbs}
\end{center}
\end{table}

After the collection of data presented here, WASA-at-COSY has collected a
high statistics data sample of $\eta$ mesons using the proton-proton production
reaction. This new data set is particularly important for rare
decay studies since an order of magnitude increase in the number of $\eta$ meson
decay events is expected. The background to signal ratio and
the detector resolution are comparable to the presented 
$pd$ data.
\begin{acknowledgments}

This work was supported in part by the EU Integrated Infrastructure Initiative HadronPhysics Project under contract number RII3-CT-2004-506078; by
the European Commission under the 7th Framework Programme through the
’Research Infrastructures’ action of the ’Capacities’ Programme, 
Call: FP7-INFRASTRUCTURES-2008-1, Grant Agreement N. 227431; by the
Polish National Science Centre through the Grant No.  0320/B/H03/2011/40, 2013/11/N/ST2/04152, 2011/03/B/ST2/01847, and the Foundation for Polish Science (MPD). We gratefully acknowledge the support given by the Swedish Research Council, the Knut and Alice Wallenberg Foundation, and the Forschungszentrum J\"ulich FFE Funding Program of the J\"ulich Center for Hadron Physics.

This work is based on the PhD theses of Daniel Coderre, Ma{\l}gorzata Hodana and Patrick Wurm. 
\end{acknowledgments}

\bibliography{biblio}







\end{document}